\documentclass[review,1p,times,authoryear]{elsarticle}

\usepackage{amssymb,amsmath,mathtools,xcolor,graphicx,xspace,colortbl,ragged2e,rotating} %
\usepackage{amsfonts}  
\usepackage{amsthm}  
\usepackage{color}  
\usepackage{hyperref}
\usepackage{enumerate}  
\providecommand{\U}[1]{\protect\rule{.1in}{.1in}}
\allowdisplaybreaks
\newtheorem{theorem}{Theorem}

\newtheorem{definition}[theorem]{Definition}
\newtheorem{example}[theorem]{Example}

\newtheorem{lemma}[theorem]{Lemma}
\newtheorem{notation}[theorem]{Notation}
\newtheorem{problem}[theorem]{Problem}
\newtheorem{proposition}[theorem]{Proposition}
\newtheorem{remark}[theorem]{Remark}

\usepackage{amsmath}

\journal{Journal of Symbolic Computation}

\begin{document}

\begin{frontmatter}
\title{Subresultants of Several Univariate Polynomials in Newton Basis\tnoteref{tl}} \tnotetext[t1]{This article is part of the volume titled ``Computational Algebra and Geometry: A special issue in memory and honor of Agnes Szanto".  }

\author[1]{Weidong Wang}
\ead{weiyang020499@163.com}

\author[1]{Jing Yang\corref{cor}}
\ead{yangjing0930@gmail.com}
\cortext[cor]{Corresponding author}
\ead[url]{https://jyangmath.github.io/}

\affiliation[1]{organization={SMS--HCIC--School of Mathematics and Physics,\\
		Center for Applied Mathematics of Guangxi, \\Guangxi Minzu University},
	addressline={\\188 Daxue East Road}, 
	city={Nanning},
	postcode={530006}, 
	state={Guangxi},
	country={China}}
\begin{abstract}
In this paper,
we consider the problem of formulating the subresultant polynomials for several univariate polynomials in Newton basis. It is required that the resulting subresultant polynomials be expressed in the same Newton basis as that used in the input polynomials. To solve the problem, we devise a particular matrix with the help of the companion matrix of a polynomial in Newton basis. Meanwhile, the concept of determinant polynomial in power basis for formulating subresultant polynomials is extended to that in Newton basis. It is proved that the generalized determinant polynomial of the specially designed matrix provides a new formula for the subresultant polynomial in Newton basis, which is equivalent to the subresultant polynomial in power basis.
Furthermore, we show an application of the new formula in devising a basis-preserving method for computing the gcd of several Newton polynomials.
\end{abstract}



\end{frontmatter}

\newpage

\section{Introduction}
\noindent Resultant theory is a fundamental tool in computer algebra with numerous applications (e.g., \citet{
	1998_Gonzalez_Recio_Lombardi,
	2008_Szanto,
	2017_Imbach_Moroz_Pouget,
	2018_Perrucci_Roy,
	2020_Roy_Szpirglas}),
such as polynomial system solving (e.g., \citet{kapur1994,wang1998,wang2000}) and quantifier elimination (e.g., \citet{arnon1984,collins1991}).
Due to its importance,
extensive research has been carried out both in theoretical and practical aspects on resultants, subresultants, and their variants (see \citet{sylvester1853,collins1967,barnett1971greatest,lascoux2003,terui2008,bostan2017,hong2021subresultant,cox2021}). One of the essential topics in resultant theory is the representation of resultants and subresultant polynomials.
In resultant theory, the input polynomials are typically assumed to be expressed in power basis (i.e., standard basis), and so are the output subresultant polynomials. However, with the rising popularity of basis-preserving algorithms in various applications (see \citet{fr1987,goodman1991,cp1993,jj2003,bl2004,aadr2006,cl2006,mm2007,acgs2007}), people are more and more interested in resultants and subresultants for polynomials in non-standard basis. Among these bases, Newton basis is widely used and has many applications. 
A typical application  is to formulate the interpolating polynomials. One may wonder whether the resulting Newton polynomials have common zeros or what their gcd looks like. These problems can be reduced to the computation of resultant/subresultants. One straightforward way is changing from Newton basis to power basis,
	then carrying out the required computations, and changing back to the initial basis afterwards. However, due to the numerical instability caused by  basis transformation,  it is desired to have a basis-preserving algorithm for computing resultant/subresultants that does not involve basis transformation. In other words, given two polynomials in Newton basis, we want to compute their gcd in the given basis, where basis transformation is not used.
In this paper, we will focus on the problem of formulating a type of subresultant polynomials for multiple univariate polynomials in Newton basis, and we aim to ensure that the formulated subresultant polynomials are: (1) expressed in the given Newton basis and (2) the same as those formulated in power basis after expansion.

Subresultant polynomials are usually expressed in the form of determinant polynomials. This choice is because determinant polynomials exhibit nice algebraic properties, greatly facilitating theoretical developments and subsequent practical applications.
Consequently, people have developed various formulas in the form of determinant polynomials for subresultant polynomials in power basis, including the Sylvester type (\citet{sylvester1853,li2006}), B\'ezout type (\citet{houwang2000}), Barnett type (\citet{barnett1983,diaz2002}), and other variants (\citet{diaz2004various}). Following this approach, we develop a determinant polynomial formula for subresultant polynomials of multiple univariate polynomials in Newton basis. 

To develop the formula of subresultant polynomial for multiple univariate polynomials in Newton basis, we use the extended version of the well-known companion matrix of a polynomial in power basis to Newton basis. This extension is achieved based on the understanding that the companion matrix of a polynomial in power basis with degree $n$ can represent an endomorphism of the spaces consisting of polynomials with degree less than $n$ defined by the multiplication of the involved variable in the power basis. The companion matrix of a polynomial in Newton basis possesses a similar property. Their only difference is that we use  Newton basis this time. This underlying idea has been conceived in \citet{diaz2002,diaz2004various}.
In \citet{diaz2002}, the authors reformulated Barnett's theorem with respect to various non-standard basis and demonstrated that the reformulated matrices partially share several nice properties with Barnett's matrices, e.g., their ranks are equal, and the indices for independent rows are the same. However, it remains unsolved how to express the gcd of the input polynomials in the given basis.
In \citet{diaz2004various}, the authors considered a specific Newton basis with nodes being the roots of one polynomial and constructed the subresultant polynomials of two polynomials. The constraint on nodes is removed in the formula presented in this paper.

With this essential concept of companion matrix extended to Newton basis, we proceed to construct a matrix that will be used to formulate the subresultant polynomials
by utilizing a similar approach as adopted by  \citet{barnett1969,barnett1970,barnett1971greatest}, \citet{diaz2004various}, and \citet{hong2021subresultant}. To express the subresultant polynomials in terms of the given Newton basis, we generalize the concept of determinant polynomial in power basis to that in Newton basis. This generalization allows the formulation of subresultant polynomials in the provided Newton basis. 
	Furthermore, as an application of the new formula, we devise a method for computing the gcd of several numerical Newton polynomials, which does not involve basis transformation.

Compared with previous related works, the newly developed formula of subresultant polynomials for multiple polynomials has the following features. First, it can be viewed as a generalization of the Barnett-type subresultant polynomials by extending the basis from power basis to Newton basis. Second, it also serves as a generalization of the formula of subresultant polynomial based on roots proposed in \citet{hoon2002} and \citet{diaz2004various} in the sense that the previous formulas use a specific Newton basis with nodes to be the roots of one of the given polynomials. In contrast, the Newton basis employed in this paper allows for an arbitrary choice of nodes. As there are infinitely many possibilities for the nodes of the Newton basis, an infinite number of formulas of subresultant polynomials can be obtained. Furthermore, the formula of subresultant polynomials developed in this paper applies to the multi-polynomial case.

The paper is structured as follows. Section \ref{sec:preliminaries} briefly reviews the subresultant theory for two polynomials. It is followed by a formal statement of the problem to be addressed by the current paper in Section \ref{sec:problem}. The main result of the paper provides a solution to the problem raised in the previous section and is elaborated in Section \ref{sec:main_result}.
The correctness of the main result is verified in Section \ref{sec:proofs}. In Section \ref{sec:application}, we show an application of the main result in computing the gcd of several numerical Newton polynomials. The paper is concluded in Section \ref{sec:conclusion} with further remarks.


\section{Preliminaries}\label{sec:preliminaries}

This section reviews some well-known results in subresultant theory for two univariate polynomials. It is noted that the involved polynomials are all expressed in power basis. Throughout the paper, we assume  $\mathbb{F}$ is the fraction field of an integral domain and $\mathbb{F}_n[x]$ is the vector space consisting of polynomials in $x$ with degree no greater than $n$.

\subsection{Sylvester subresultant polynomial of two polynomials}
Subresultant polynomials for two univariate polynomials are usually defined with the minors of their Sylvester matrix. Thus we start by recalling the definition of Sylvester matrix below.

Consider $A,B\in\mathbb{F}[x]$ with the following form:
\begin{align}
	A(x) &= a_nx^n+\cdots+a_1x+a_0, \label{eq:A_coef}\\
	B(x) &= b_mx^m+\cdots+b_1x+b_0 \label{eq:B_coef},
\end{align}
where $a_nb_m\ne0$, $a_i,b_j\in\mathbb{F}$~($0\le i\le n, 0\le j\le m$).
Then
the Sylvester matrix of $A$ and $B$ with respect to $x$
is defined as$${\rm{Syl}}(A,B) := \underbrace{\left[\begin{array}{*{20}{l}}
		{{a_n}}& \cdots &{{a_0}}&{}\\
		& \ddots &  & \ddots &{}\\
		&{}&{{a_n}}& \cdots &{{a_0}}\\\hline
		{{b_m}}& \cdots &{{b_0}}&{}\\
		& \ddots & & \ddots &{}\\
		&{}&{{b_m}}& \cdots &{{b_0}}
	\end{array}\right]}_{(m+n)\ \text{columns}}\hspace{-1.8em}\begin{array}{*{20}{l}}
	{\left. {\begin{array}{*{20}{c}}
				{}\\[1pt]
				{}\\[1pt]
				{}
		\end{array}} \right\}}{}~m\ \text{rows}\\[15pt]
	{\left. {\begin{array}{*{20}{c}}
				{}\\[1pt]
				{}\\[1pt]
				{}
		\end{array}} \right\}}{}~n\ \text{~rows}
\end{array}$$

To define the subresultant polynomial of $A$ and $B$, we introduce the concept of determinant polynomial (e.g., see \citet[Definition 7.5.1]{1993_Mishra}). 
\begin{definition}[Determinant polynomial]\label{def:detp_power}
	Given $M\in \mathbb{F}^{(n-k)\times n}$, the determinant polynomial of $M$ in terms of $x$ is defined as $$\operatorname*{detp}M:=\sum_{i=0}^{k}\det M_i\cdot x^{i},$$
	where $M_i$ is the submatrix of $M$ consisting of
	the first $n-k-1$ columns and the $(n-i)$th column.
\end{definition}

With the above settings, the definition of subresultant polynomials of two
univariate polynomials is presented below. 

\begin{definition}[Subresultant polynomial, \citet{diaz2004various}]
	For $0\le k\le\min(m,n)-1$, the $k$th subresultant polynomial of $A$ and $B$ with respect to $x$ is defined as
	$$S_k(A,B):=\operatorname*{detp}{\rm Syl}_k(A,B),$$
	where ${\rm Syl}_k(A,B)$
	is the submatrix of ${\rm Syl}(A,B)$ obtained by deleting the last $k$ rows from the upper block (consisting of the first $m$ rows) and the lower block (consisting of the last $n$ rows) respectively and the last $k$ columns.
\end{definition}

\begin{example}\label{ex:Sk_AB}
	Given
	$$A(x)=4x^3 - 8x^2 + 5x - 1, \quad B(x)=2x^3 +3 x^2 - 1,$$
	we have
	$${\rm Syl}_1(A,B)=\begin{bmatrix}
		4&-8&\ \ 5&-1&\\
		&\ \ 4&-8&\ \ 5&-1\\
		2&\ \ 3&\ \ 0&-1&\\
		&\ \ 2&\ \ 3&\ \ 0& -1\\
	\end{bmatrix}$$
	and the first subresultant polynomial of $A$ and $B$ with respect to $x$ is
	\begin{align*}
		S_1(A,B)&={\rm detp}\
		{\rm Syl}_1(A,B)\\
		&=\det \begin{bmatrix}
			4&-8&\ \ 5&-1\\
			&\ \ 4&-8&\ \ 5\\
			2&\ \ 3&\ \ 0&-1\\
			&\ \ 2&\ \ 3&\ \ 0\\
		\end{bmatrix}x+\det
		\begin{bmatrix}
			4&-8&\ \ 5&\\
			&\ \ 4&-8&-1\\
			2&\ \ 3&\ \ 0&\\
			&\ \ 2&\ \ 3&\  -1\\
		\end{bmatrix}\\
		&=576\,x-288.
	\end{align*}
\end{example}

Apart from Sylvester subresultant polynomials, people also developed various alternative expressions for subresultant polynomials. Depending on the types of matrices used to formulate them, we can categorize the expressions into Sylvester type, B\'ezout type,  Barnett type, etc. (see \citet{diaz2004various} for more details).

\subsection{Companion matrix  and Barnett matrix}
Companion matrix has been proved to be a useful tool for formulating subresultant polynomials of two univariate polynomials (see \citet{diaz2004various}). We give a brief review of the companion matrix below.

Given a polynomial $A(x)=a_nx^n+a_{n-1}x^{n-1}+\cdots+a_0\in\mathbb{F}[x]$, the companion matrix of $A(x)$ is defined as an $n\times n$ matrix $\Delta_A$ of the following form
\[{\Delta _A} = \left[ {\begin{array}{*{20}{c}}
		0&0& \cdots &0&{ - {a_0}}\\
		a_{n}&0& \cdots &0&{ - {a_1}}\\
		0&a_{n}& \cdots &0&{ - {a_2}}\\
		\vdots & \vdots & \ddots & \vdots & \vdots \\
		0&0& \cdots &a_{n}&{ - {a_{n - 1}}}
\end{array}} \right]_{n\times n}\]

With the help of the companion matrix above, we may construct the Barnett matrix of two univariate polynomials, whose minors give the coefficients of subresultant polynomials.

More explicitly,
let $A$ and $B$ be as in \eqref{eq:A_coef} and \eqref{eq:B_coef}, respectively.
Then $B(\Delta _A/a_n)$ is the Barnett matrix of $A$ and $B$ with respect to $x$. Furthermore, let $B_k$ be the submatrix of  $B(\Delta_A/a_n)$ obtained by deleting the last $k$ columns and $B_{ik}$ be the submatrix of $B_k$ consisting of the last $n-k-1$ rows and the $(i+1)$th row. Then by \citet[Theorem 2.2-1]{diaz2004various},
\[S_{k}(A,B)=a_n^{m-k}\sum_{i=0}^k\det B_{ik}\cdot x^i\]

\begin{example}
	Consider the polynomials $A$ and $B$ in Example \ref{ex:Sk_AB}. Then
	with some calculations, we obtain
	\[\Delta_A=
	\left[ \begin {array}{rrr}
	0&0&1\\
	4&0&-5\\
	0&4&8
	\end {array} \right],
	\quad B(\Delta_A/a_n)= \left[
	\begin{array}{rrr}
		-{\dfrac{1}{2}}&{\dfrac{7}{4}}&{\dfrac{23}{8}}\\[8pt]
		-{\dfrac{5}{2}}&-{\dfrac{37}{4}}&-{\dfrac{101}{8}}
		\\[8pt]
		7&{\dfrac{23}{2}}&{\dfrac{55}{4}}
		\end {array}
		\right]
		\]
		It follows that
		\[S_1(A,B)=4^{3-1}\left(\det\left[\begin{array}{rrr}-{\dfrac{5}{2}}&-{\dfrac{37}{4}}\\[8pt]
			{7}&{\dfrac{23}{2}}\end{array}\right]x+
		\det\left[\begin{array}{rrr}
			-{\dfrac{1}{2}}&{\dfrac{7}{4}}\\[8pt]
			{7}&{\dfrac{23}{2}}
		\end{array}\right]\right)=576\,x-288\]
	\end{example}
	
	\subsection{Subresultant polynomial in roots}
	The subresultant polynomials in the previous subsections are all expressed in coefficients. In \citet{1999_Hong,hoon2002} and \citet{2006D'Andrea}, Hong et al. developed an equivalent formula in roots for subresultant polynomials, which inspires the authors in \citet{hong2021subresultant} with a promising way to extend the concept of subresultant polynomial from two polynomials to multiple polynomials.
	
	Given $A,B\in\mathbb{F}[x]$ with degrees $n$ and $m$, respectively, let $a_n$ be the leading coefficient of $A$ and $\alpha_1,\ldots,\alpha_n$
	be the $n$ roots of $A$ over the algebraic closure of~$\mathbb{F}$. Then \begin{equation}\label{eq:Sk_AB_roots}
		S_k(A,B)=\dfrac{a_{n}^{m-k}\cdot
				\det\left[
				\begin{array}{rcr}
					\alpha_1^{n-k-1}B(\alpha_1)&\cdots&\alpha_n^{n-k-1}B(\alpha_n)\\
					\vdots~~~~~~~~&&\vdots~~~~~~~~\\
					\alpha_1^{0}B(\alpha_1)&\cdots&\alpha_n^0B(\alpha_n)\\\hline
					\alpha_1^{k-1}(x-\alpha_1)&\cdots&\alpha_n^{k-1}(x-\alpha_n)\\
					\vdots~~~~~~~~&&\vdots~~~~~~~~\\
					\alpha_1^0(x-\alpha_1)&\cdots&\alpha_n^0(x-\alpha_n)
				\end{array}
				\right]_{n\times n}}{\det
				\begin{bmatrix}
					\alpha_1^{n-1}&\cdots&\alpha_n^{n-1}\\
					\vdots&&\vdots\\
					\alpha_1^0&\cdots&\alpha_n^{0}\end{bmatrix}_{n\times n}}
	\end{equation}

	The above rational expression for $S_{k}$ in roots should be interpreted as follows; otherwise, the denominator will vanish when $A$ is not squarefree.
	\begin{enumerate}[(1)]
		\item Treat\ $\alpha_1,\ldots,\alpha_{n}$ as indeterminates and carry out the exact division, which results in a symmetric polynomial in terms of $\alpha_1,\ldots,\alpha_{n}$.
		\item Evaluate the obtained polynomial with $\alpha_1,\ldots,\alpha_{n}$ assigned the values of roots of $A$.
	\end{enumerate}
	
	\begin{example}
		Consider the polynomials $A$ and $B$ in Example \ref{ex:Sk_AB}. It is noted that the roots of $A$ are $(\alpha_1,\alpha_2,\alpha_3)=(1/2,1/2,1)$. With further calculation, we have
		\begin{align*}
			S_1(A,B)=&\dfrac{ 4^{3-1}\cdot
				\det\begin{bmatrix}
					\alpha_1^1B(\alpha_1)&\alpha_2^1B(\alpha_2)&\alpha_3^1B(\alpha_3)\\
					\alpha_1^0B(\alpha_1)&\alpha_2^0B(\alpha_2)&\alpha_3^0B(\alpha_3)\\
					\alpha_1^0(x-\alpha_1)&\alpha_2^0(x-\alpha_2)&\alpha_3^0(x-\alpha_3)
			\end{bmatrix}}{\det
				\begin{bmatrix}
					\alpha_1^2&\alpha_2^2&\alpha_3^2\\
					\alpha_1^1&\alpha_2^1&\alpha_3^1\\
					\alpha_1^0&\alpha_2^0&\alpha_3^0\end{bmatrix}}
			\\
			=\,& 16\big((\,{4\alpha_{{1}}^{2}}{\alpha_{{2}}^{2}}+4\,{\alpha_{{1}}^{2}}{\alpha_{{3}}^{2}}+4\,{\alpha_{{2}}^{2}}{\alpha_{{3}}^{2}}+4\,{\alpha_{{1}}^{2}}\alpha_{{
					2}}\alpha_{{3}}+4\,\alpha_{{1}
			}{\alpha_{{2}}^{2}}\alpha_{{3}}+4\,\alpha_{{1}}\alpha_{{2}}{\alpha_{{3
				}}^{2}}+6\,{\alpha_{{1}}^{2}}
			\alpha_{{2}}\\
			&+6\,{\alpha_{{1}}^{2}}\alpha_{{3}}+6\,\alpha_{{1}}{\alpha_
				{{2}}^{2}}+6\,\alpha_{{1}}{
				\alpha_{{3}}^{2}}+6\,{\alpha_{{2}}^{2}}\alpha_{{3}}+6\,\alpha_{{2}}{
				\alpha_{{3}}^{2}}+12\,\alpha_{{1}}\alpha_{{2}}\alpha_{{3}}+9\,\alpha_{{1}}\alpha_{{2}}+9\,\alpha_{{1}}\alpha_{{
					3}}\\
			&+9\,\alpha_{{2}}\alpha_{{3}}+2\,\alpha_{{1}}+2\,\alpha_{{2}}+2\,
			\alpha_{{3}}+3)x-(4\,{\alpha_{{1}}^{2}}{\alpha_{{2}}^{2}}\alpha_{{3}}+4\,{\alpha_{{1}}^{2}}
			\alpha_{{2}}{\alpha_{{3}}^{2}}+4\,\alpha_{{1}}{\alpha_{{2}}^{2}}{
				\alpha_{{3}}^{2}}\\
			&+6\,{\alpha_{{1}}^{2}}\alpha_{{2}}\alpha_{{3}}+6\,
			\alpha_{{1}}{\alpha_{{2}}^{2}}\alpha_{{3}}+6\,\alpha_{{1}}\alpha_{{2}}
			{\alpha_{{3}}^{2}}+9\,\alpha_{{1}}\alpha_{{2}}\alpha_{{3}}+2\,{\alpha_
				{{1}}^{2}}+2\,
			{\alpha_{{2}}^{2}}+2\,{\alpha_{{3}}^{2}}\\
			&+2\,\alpha_{{1}}\alpha_{{2}}+2\,\alpha_{{1}}\alpha_{{3}}+2\,\alpha_{{2}}\alpha_{{3}}+3
			\,\alpha_{{1}}+3\,\alpha_{{2}}+3\,\alpha_{{3}})
			\big)
		\end{align*}
		The evaluation of $S_{1}(A,B)$ with $(\alpha_1,\alpha_2,\alpha_3)=(1/2,1/2,1)$ yields $S_{1}(A,B)=576\,x-288.$
	\end{example}

	\section{Problem Statement}\label{sec:problem}
	
	In this section, we formalize the problem to be addressed in the current paper. For this purpose, we
	begin with a review of the generalized subresultant polynomial of multiple univariate polynomials, developed in recent years by \citet{hong2021subresultant}. 

	\subsection{Subresultant polynomial of several univariate polynomials}
	We need the following notations to present the generalized subresultant polynomial
	for more than two polynomials.
	\begin{notation}\label{notation} \
		\begin{itemize}
			\item $F = ({F_0},{F_1}, \ldots ,{F_t})\subseteq \mathbb{F}[x]$;
			\item ${d_i} = \deg {F_i}$;
			\item $\alpha_1,\ldots,\alpha_{d_0}$ are the $d_0$ roots of $F_0$  in the algebraic closure of $\mathbb{F}$;
			\item
			$\delta  = ({\delta _1},{\delta _2}, \ldots ,{\delta _t})\in \mathbb{N}^t$;\footnote{It is assumed that $0\in\mathbb{N}$.}
			\item $\left| \delta  \right| = {\delta _1} +   \cdots  + {\delta _t} \le {d_0}$.
		\end{itemize}
	\end{notation}
	
	With the above notations, we recall the definition of the $\delta$th subresultant polynomial for several univariate polynomials in terms of roots from \citet[Definition 2 and Lemma 30]{hong2021subresultant}, which can be viewed as the generalization of Equation \eqref{eq:Sk_AB_roots} to multiple polynomials.
	We point out that the expression of subresultant polynomial we use in the current paper differs from that in \citet{hong2021subresultant} by a sign factor. The reason for the sign adjustment is that we want to make the sign match with the equivalent form of subresultant polynomials in coefficients as developed in a more recent work by \citet{2023_Hong_Yang}, which naturally extends the concept of Sylvester subresultant polynomial from two polynomials to several polynomials.
	
	\begin{definition}\label{def:sres}
		The $\delta$th subresultant polynomial $S_{\delta}$ of $F$ is defined by
		\begin{equation}\label{eqs:sres}
			{S_\delta }(F): = a_{0d_0}^{\delta_0}\cdot\det M_{\delta}/\det V
		\end{equation}
		where $a_{0d_0}$ is the leading coefficient of $F_0$ in terms of $x$, and
		\begin{itemize}
			\item $M_{\delta}=\left[ {
					\begin{array}{rcr}
						{\alpha _1^{{\delta _1} - 1}{F_1}({\alpha _1})}& \cdots &{\alpha _{{d_0}}^{{\delta _1} - 1}{F_1}({\alpha _{{d_0}}})}\\
						\vdots~~~~~~~~ &  & \vdots~~~~~~~~ \\
						{\alpha _1^{0}{F_1}({\alpha _1})}& \cdots &{\alpha _{{d_0}}^{0}{F_1}({\alpha _{{d_0}}})}\\
						\hline
						\vdots~~~~~~~~ &{}& \vdots~~~~~~~~ \\
						\vdots~~~~~~~~ &{}& \vdots~~~~~~~~ \\
						\hline
						{\alpha _1^{{\delta _t} - 1}{F_t}({\alpha _1})}& \cdots &{\alpha _{{d_0}}^{{\delta _t} - 1}{F_t}({\alpha _{{d_0}}})}\\
						\vdots~~~~~~~~ &  & \vdots~~~~~~~~ \\
						{\alpha _1^{0}{F_t}({\alpha _1})}& \cdots &{\alpha _{{d_0}}^{0}{F_t}({\alpha _{{d_0}}})}\\
						\hline
						{\alpha _1^{\varepsilon  - 1}(x - {\alpha _1})}& \cdots &{\alpha _{{d_0}}^{\varepsilon  - 1}(x - {\alpha _{{d_0}}})}\\
						\vdots~~~~~~~~ &  & \vdots~~~~~~~~ \\
						{\alpha _1^{0}(x - {\alpha _1})}& \cdots &{\alpha _{{d_0}}^{0}(x - {\alpha _{{d_0}}})}
				\end{array}} \right]_{d_0\times d_0}$
			\item $V = \left[ {\begin{array}{*{20}{c}}
						{\alpha _1^{{d_0} - 1}}& \cdots &{\alpha _{{d_0}}^{{d_0} - 1}}\\
						\vdots &{}& \vdots~ \\
						{\alpha _1^{0}}& \cdots &{\alpha _{d_0}^{0}}
				\end{array}} \right]_{d_0\times d_0}$
			\item $\delta_0=\max(d_1+\delta_1-d_0,\ldots,d_t+\delta_t-d_0,1-|\delta|);$
			\item $\varepsilon  = {d_0} - |\delta |$.
		\end{itemize}
		The coefficient of $S_{\delta}(F)$ in the term $x^{\varepsilon}$ is called the {leading coefficient} of $S_{\delta}(F)$, denoted by $s_{\delta}(F)$. If no ambiguity occurs, we can write $S_{\delta}(F)$ as $S_{\delta}$ for simplicity.
	\end{definition}
	
	\begin{remark}\
		\begin{enumerate}[(1)]
			\item It is emphasized that the rational expression \eqref{eqs:sres} in Definition \ref{def:sres} should be interpreted the same way as that for interpreting Equation \eqref{eq:Sk_AB_roots}.
			\item When $\delta_i=0$ for some $i>0$, the block of $M_{\delta}$ involving $F_i$ will disappear and $S_{\delta}$ will not contain the information of $F_i$.
			\item It is noted that $\varepsilon$ captures the formal degree of $S_{\delta}$ in terms of $x$.
		\end{enumerate}
	\end{remark}

	The choice for $\delta_0$ in Definition \ref{def:sres} seems artificial. Now we justify why $\delta_0$ is chosen in this particular way. The reason is to make $S_{\delta}$ be a polynomial in terms of the coefficients of $F_i$'s with the smallest degree. A more detailed explanation is given below. Again, treating $\alpha_1,\ldots,\alpha_{d_0}$ as indeterminates and carrying out the exact division for $\det M_{\delta}/\det V$, we obtain a symmetric polynomial in $\alpha_1,\ldots,\alpha_{d_0}$ which can always be converted into a polynomial in elementary symmetric polynomials (denoted by $e_1,\ldots,e_{d_0}$) in terms of $\alpha_1,\ldots,\alpha_{d_0}$. By Vieta formulas, $e_i~(1\le i\le d_0)$ is a rational function in the coefficients of $F_0$ with the denominator to be $a_{0d_0}$. To make $S_{\delta}$ a polynomial in terms the coefficients of $F_i$'s, we need to cancel the denominators by multiplying $\det M_{\delta}/\det V$ with $a_{0d_0}^p$ where $p$ is the total degree of $\det M_{\delta}/\det V$ (viewed as polynomial in $e_1,\ldots,e_{d_0}$) in terms of $e_i$'s. Next, we will figure out what $p$ is.
	Note that $e_i$ is linear in $\alpha_1$. Thus the total degree of  $\det M_{\delta}/\det V$ in terms of $e_i$'s is equal to its degree in terms of $\alpha_1$. Combining the observations
	that the degree of $\det M_{\delta}$ in  $\alpha_1$ is $\max(d_1+\delta_1,\ldots,d_t+\delta_t,\varepsilon+1)$ and that of $\det V$ is $d_0$, we have
	\begin{align*}
		p&=\max(d_1+\delta_1,\ldots,d_t+\delta_t,\varepsilon+1)-d_{0}\\
		&=\max(d_1+\delta_1-d_0,\ldots,d_t+\delta_t-d_0,1-|\delta|),
	\end{align*}
	which is exactly the value of $\delta_0$ in Definition \ref{def:sres}.
	
	\begin{example}\label{ex:3poly_pb}
		Given
		$$F = ({F_0},{F_1},{F_2}) = (4x^3 - 8x^2 + 5x - 1,2x^3 - 3x^2 + x,2x^3 + x^2 - x)$$
		and $\delta=(1,1)$,
		obviously, we have $(d_0,d_1,d_2)=(3,3,3)$.
		Moreover, it is easy to obtain by calculation that $(\alpha_1,\alpha_2,\alpha_3)=(1/2,1/2,1)$ and $$\delta_0=\max(3+1-3,3+1-3,1-(1+1))=1.$$Now
		we construct
		\begin{align*}
			V=&\begin{bmatrix}
				\alpha_1^2&\alpha_2^2&\alpha_3^2\\
				\alpha_1^1&\alpha_2^1&\alpha_3^1\\
				\alpha_1^0&\alpha_2^0&\alpha_3^0
			\end{bmatrix},\\[3pt]
			M_{\delta}=&\left[
				\begin{array}{lll}
					\alpha_1^0(2{\alpha_1^3} -3{\alpha_1^2} + \alpha_1 )&\alpha_2^0(2{\alpha_2^3} - 3{\alpha_2^2} + \alpha_2 )&\alpha_3^0(2{\alpha_3^3} - 3{\alpha_3^2} + \alpha_3 )\\
					\alpha_1^0(2{\alpha_1^3} +{\alpha_1^2} - \alpha_1 )&\alpha_2^0(2{\alpha_2^3} +{\alpha_2^2} -\alpha_2 )&\alpha_3^0(2{\alpha_3^3} +{\alpha_3^2} -\alpha_3)\\
					\alpha_1^0(x- \alpha_1)&\alpha_2^0(x- \alpha_2)&\alpha_3^0(x- \alpha_3)
				\end{array}
				\right].
		\end{align*}
		Further calculation yields
		\begin{align*}
			\det V=\,&\left( \alpha_{{2}}-\alpha_{{3}} \right)    \left( \alpha_{{1}}-\alpha_{{3}} \right)
			\left( \alpha_{{1}}-\alpha_{{2}} \right),\\
			\det M_{\delta}=\,&2\det V\cdot\left(\left( 4e_{2}-2\,e_1+1 \right) x-4\,e_3
			\right),
		\end{align*}
		where $e_1=\alpha_{{1}}+\alpha_{{2}}+\alpha_{{3
		}}$, $e_2=\alpha_{{1}}\alpha_{{2}}+\alpha_{{1}}\alpha_{{3}}+
		\alpha_{{2}}\alpha_{{3}}$, $e_3=\alpha_{{1}}\alpha_{{2}}\alpha_{{3}}$.
		Therefore, $$S_{\delta}(F)=4^{1}\cdot2(\left( 4e_{2}-2\,e_1+1 \right) x-4\,e_3
		).$$
		The evaluation of $S_{\delta}(F)$ with $(\alpha_1,\alpha_2,\alpha_3)=(1/2,1/2,1)$ yields $S_{\delta}(F)=16x-8.$
	\end{example}
	
	The $\delta$th subresultant polynomial provides a practical tool for tackling the problems of parametric gcds and parametric multiplicities because of the inherent connection it has with the incremental gcds of several univariate polynomials.
	\begin{definition}[\citet{hong2021subresultant}]
		Given $F\subseteq\mathbb{F}[x]$, let $$\theta_i=\deg\gcd(F_0,\ldots,F_{i-1})-\deg\gcd(F_0,\ldots,F_{i})\quad \text{for}\quad i=1,\ldots,t$$
		where $\gcd (F_0):=F_0$. Then $\operatorname*{icdeg}(F):=(\theta_1,\ldots,\theta_t)$ is called the incremental cofactor degree of $F$.
	\end{definition}
	
	\begin{theorem}[\citet{hong2021subresultant}]\label{thm:icdeg}
		Let $\theta=\max\limits_{s_{\delta}(F)\ne0}\delta$
		where $\delta$'s are as specified in Notation \ref{notation} and $\max$ is with respect to the ordering $\succ_{\operatorname*{glex}}$. Then we have
		$$\operatorname*{icdeg}(F)=\theta\ \ \
		\text{and}\ \ \ \gcd(F)=S_{\theta}(F).$$
	\end{theorem}
	\begin{remark}
		The ordering $\succ_{\rm{glex}}$ is defined for two sequences, e.g.,
		$\gamma$ and $\delta$, in $\mathbb{N}^{t}$. We say $\delta
		\succ_{\rm{glex}}\gamma$ if $|\delta|>|\gamma|$, or $|\delta|=|\gamma|$ and there exists $i\le t$ such that $\delta
		_{i}>\gamma_{i}$ and $\delta_{j}=\gamma_{j}$ for $j<i$.
	\end{remark}
	
	\begin{example}
		Consider $F=(F_0,F_1,F_2)$ where $F_i$'s are as in Example \ref{ex:3poly_pb}. All the possible $\delta$'s for the specific $F$ form the following set:
		\[\{\ (3,0),\ (2,1),\ (1,2),\ (0,3),\ (2,0),\ (1,1),\ (0,2),\ (1,0),\ (0,1),\ (0,0)\ \}\]
		where the entries are ordered with respect to $\succ_{\rm glex}$.
		With some calculations, we obtain
		\[s_{(3,0)}(F)=s_{(2,1)}(F)=s_{(1,2)}(F)=s_{(0,3)}(F)=s_{(2,0)}(F)=0\]
		and $s_{(1,1)}(F)\ne0$. Hence, $\gcd F=S_{(1,1)}(F)$.
	\end{example}

	\subsection{Determinant polynomial in Newton basis}
	In order to state the problem, we need to generalize the concept of determinant polynomial from power basis to Newton basis.
	
	\begin{definition}[Newton basis]\label{def:nwbasis}
		Let $\lambda=(\lambda_1,\ldots,\lambda_{n})\in\mathbb{F}^{n}$ and
		$B^{\lambda}(x) = ( {B_0},{B_1}, \ldots ,{B_{n }})$
		where
		\begin{equation}\label{eqs:nw_basis}
			B_i=(x-\lambda_1)\cdots(x-\lambda_i)
		\end{equation}
		with the convention $B_0:=1$.
		We call $B^{\lambda}(x)$\footnote[1]{$B^{\lambda}(x)$ can be abbreviated as $B^{\lambda}$ if no ambiguity occurs.} the \emph{Newton basis} of $\mathbb{F}_{n}[x]$ with respect to $\lambda$.
	\end{definition}
	
	It should be pointed out that the classical definition of Newton basis requires that $\lambda_i$'s are distinct. In contrast, we discard this restriction in this paper because the logical correctness of the result does not rely on the restriction.
	
	The following concept extends the determinant polynomial in power basis to an arbitrary polynomial set, which also covers Newton basis. The more general concept is introduced because it will play a certain role in the proofs later.
	\begin{definition}[Determinant polynomial with respect to a polynomial set]\label{def:detp}
		Let $M\in\mathbb{F}^{{(n-k)\times n}}$ where $k\le n$ and $P=(P_0,\ldots,{P_k})\subseteq\mathbb{F}[x]$.
		Then the determinant polynomial of $M$ with respect to $P$ is defined as
		\[{\rm detp}_{P}M=\sum_{i=0}^{k}\det M_i\cdot P_{i}\]
		where $M_i$ is the submatrix of $M$ consisting of
		the last $n-k-1$ columns and the $i$th column.
	\end{definition}
	
	\begin{remark}\
		\begin{itemize}
			\item When $P=(B_0,\ldots,B_k)$, Definition \ref{def:detp} gives the formal definition of determinant polynomial with respect to Newton basis.
			\item The selection of columns for generating $M_i$ is slightly different from the usual way in Definition \ref{def:detp_power}. We use the current version because it is more friendly for the statement of the main result.
		\end{itemize}
	\end{remark}

	\subsection{Problem statement}
	Now we are ready to give a formal statement of the problem to be solved in the paper.
	\begin{problem}\label{prob:statement}\
		\begin{description}
			\item[Given:] $F=(F_0,\ldots,F_t)\subseteq\mathbb{F}[x]$ where $F_i$'s are expressed in the Newton basis~$B^{\lambda}=(B_0,B_1,\ldots)$ with respect to $\lambda$, and $\delta\in\mathbb{N}^{t}$ such that $|\delta|\le d_0$ where $d_0=\deg F_0$
			\item[Find:] a matrix $N_{\lambda,\delta}(F)$ such that $S_\delta(F)=c\cdot{\rm detp}_{(B_0,\ldots,B_{\varepsilon})}N_{\lambda,\delta}(F)$ for some constant $c$ where $\varepsilon=d_0-|\delta|$.
		\end{description}
	\end{problem}
	
	\section{Main Result}\label{sec:main_result}
	
	This section is devoted to presenting the main result of the paper, i.e., a particular solution to Problem \ref{prob:statement}. In order to construct the desired subresultant matrix, we introduce the companion matrix of a polynomial in Newton basis.
	
	\begin{definition}[Companion matrix in Newton basis, \citet{2001_Corless_Litt}]\label{def:general-com}
		Let $B^{{\lambda}}=(B_0,\ldots,B_{n})$ be the Newton basis of $\mathbb{F}_{n}[x]$ with respect to $\lambda=(\lambda_1,\ldots,\lambda_{n})\in \mathbb{F}^{n}$.
		Let $P = {p_n}{B_n} +  \cdots  + {p_1}B_1 + {p_{0}B_0}\in\mathbb{F}[x]$ where $p_n\ne0$. Then the companion matrix of $P$ in $B^{{\lambda}}$, denoted by $\Lambda _{\lambda,P}$, is
		defined as$${\Lambda _{\lambda,P}} = p_n\left[ {\begin{array}{*{20}{c}}
				{{\lambda _1}}&{}&{}&{}&{ - {p_{0}}/{p_n}}\\
				1&{{\lambda _2}}&{}&{}&{ - {p_{1}}/{p_n}}\\
				{}&1& \ddots &{}& \vdots \\
				{}&{}& \ddots &{{\lambda _{n - 1}}}&{- {p_{n-2}}/{p_n}}\\
				{}&{}&{}&1&{\lambda_n- {p_{n-1}}/{p_n}}
		\end{array}} \right]_{n\times n}$$
	\end{definition}
	
	Similar to the companion matrix of a polynomial in power basis,
	the companion matrix ${\Lambda _{\lambda,P}}$  defines
	an endomorphism
	of $\mathbb{F}_{n-1}[x]$ given by the multiplication by $p_{n}x$ with respect to the Newton basis (as detailed by Proposition \ref{the:general-companion}-(1) in Section \ref{sec:proofs}).
	
	Now we are ready to state the main result of this paper.
	
	\begin{theorem}[Main result]\label{thm:sres}
		When $\delta\ne(0,\ldots,0)$, a solution to Problem \ref{prob:statement} is
		\begin{equation}\label{eqs:N_lambda_delta}
			{N}_{\lambda,\delta}(F) =
			\begin{bmatrix}
				{R_{11}}&\cdots&R_{1\delta_1}&\cdots&\cdots
				&{R_{t1}}&\cdots&R_{t\delta_t}
			\end{bmatrix}^T
		\end{equation}
		with $c=(-1)^{\sigma}\cdot a_{0d_0}^{\delta_0}$,
		where
		\begin{itemize}
			\item $R_{ij}$ is the $j$th column of $F_i({\Lambda _{{\lambda,F_0}}}/a_{0d_0})$,
			\item  ${\Lambda _{{\lambda,F_0}}}$ is the companion matrix of $F_0$ in the basis $(B_0,\ldots,B_{d_0})$,
			\item $\sigma=\left(\sum_{i=1}^{t}\binom{\delta_i}{2}\right)+\binom{\varepsilon}{2}-\binom{d_0}{2}+(d_0-1)\varepsilon$,\footnote[1]{When $m< p$, $\binom{m}{p}:=0$.} and
			\item $d_0$, $a_{0d_0}$, $\varepsilon$ and $\delta_0$ are as in Definition \ref{def:sres}.
		\end{itemize}
		Equivalently,
		$${S_\delta }(F)= (-1)^{\sigma}\cdot a_{0d_0}^{\delta_0}\cdot\operatorname*{detp}\nolimits_{(B_0,\ldots,B_{\varepsilon})} {N}_{\lambda,\delta}(F)$$
	\end{theorem}
	
	We make a few observations on Theorem \ref{thm:sres}.
	\begin{remark}\
		\begin{enumerate}[(1)]
			\item Note that $N_{\lambda,\delta}$  is of order  $|\delta|\times d_0$. If $\delta_i=0$, the block from $F_i({\Lambda _{{\lambda,F_0}}}/a_{0d_0})^T$ does not appear in ${{N}_{\lambda,\delta} }(F)$.
			
			\item When $\delta=(0,\ldots,0)$, the matrix ${{N}_{\lambda,\delta} }(F)$ is null, and thus it cannot provide any information about the $\delta$th subresultant polynomial. On the other hand, by Definition \ref{def:sres}, we have
			\[S_{(0,\ldots,0)}(F)=a_{0d_0}^{\delta_0-1}F_0.\]
			Therefore, we assume that $\delta\ne(0,\ldots,0)$ in the rest of the paper.
			
			\item From the construction of $N_{\lambda,\delta}(F)$, it is seen that $F_1,\ldots,F_t$ are not necessarily to be expressed in Newton basis.
			
			\item When $F$ is specialized with $(A,B)$ where $\deg A=n$ and $\lambda$ is specialized with $(\beta_1,\ldots,\beta_{n})$ which are the roots of $A$, ${S_{(n-k)}}$ derived from Theorem \ref{thm:sres} is equivalent to $S_{k}$ derived from the formula in \citet[Proposition 5.1]{diaz2004various} and \citet{1999_Hong}. Both can be viewed as subresultant polynomials in terms of the
			roots of the given polynomials.
		\end{enumerate}
	\end{remark}

	\begin{example}\label{ex:sres_multi_polys}
		Consider
		$$F = ({F_0},{F_1},{F_2}) = (4B_{3}-8B_{2}+9B_{1},2B_{3}-3B_{2}+3B_{1},2B_3-B_2-B_{1}+2B_{0})$$
		and $\delta=(1,1)$, where
		$$B^{\lambda}=(B_0,B_1,B_2,B_3)=(1,\  x-1,\ (x-1)(x+1),\ x(x-1)(x+1))$$
		and $\lambda=(1,-1,0)$. It is easy to verify that when converted into expressions in power basis, $F_i$'s are the same as those in Example  \ref{ex:3poly_pb}. We construct the companion matrix of $F_0$ in $B^{\lambda}$ and obtain
		\[\Lambda_{\lambda,F_0}=4
		\begin{bmatrix}
			1&\ \ 0&\ \ 0\\
			1&-1&-\dfrac{9}{4}\\[5pt]
			0&\ \ 1&\ \ 2
		\end{bmatrix}\]
		It follows that
		\[F_1(\Lambda_{\lambda,F_0}/4)=\left[ \begin {array}{ccc} \ \ 0&\ \ 0&\ \ 0\\ [2pt]
		-{\dfrac{3}{2}}&{
			\ \ \dfrac{3}{4}}&\ \ {\dfrac{9}{8}}\\[7pt]
		\ \ 1&-{\dfrac{1}{2}}&-{\dfrac{3}{4}}
		\end {array} \right]\quad\text{and}\quad F_2(\Lambda_{\lambda,F_0}/4)=
		\left[ \begin {array}{ccc}
		\ \ 2&\ \ 0&\ \ 0\\[2pt]
		-{\dfrac{7}{2}}&-
		{\dfrac{9}{4}}&-{\dfrac{27}{8}}\\[7pt]
		\ \ 5&\ \ {\dfrac{3}{2}}&{
			\ \ \dfrac{9}{4}}\end {array} \right]
		\]
	
			Given  $\delta=(1,1)$, by Theorem \ref{thm:sres}, 
			\begin{align*}
				{N}_{\lambda,\delta}(F) =&
				\begin{bmatrix}
					\ \ 0&\ \ 2\\[2pt]
					-{\dfrac{3}{2}}&{
						-\dfrac{7}{2}}\\[7pt]
					\ \ 1&\ \ 5
				\end{bmatrix}^T=
				\begin{bmatrix}
					0&-\dfrac{3}{2}&1\\[7pt]
					2&-\dfrac{7}{2}&5
				\end{bmatrix}
			\end{align*}
			Hence
			\begin{align*}
				\varepsilon=\,&3-(1+1)=1\\
				\sigma=\,&\binom{1}{2}+\binom{1}{2}+\binom{1}{2}-\binom{3}{2}+(3-1)\cdot 1=-1\\[3pt]
				\delta_0=&\max(3+1-3,3+1-3,1-(1+1))=1\\
				S_{\delta}(F)=\,&c\cdot{\rm detp}_{(B_0,B_1)}{N}_{\lambda,\delta}(F)=(-1)^{-1}\cdot4^1\cdot(-4B_1-2B_{0})=16x-8
		\end{align*}
	\end{example}
	
	\section{Proof of the Main Result (Theorem \ref{thm:sres})}\label{sec:proofs}
	For readers to better understand the proof details, we give a brief sketch of ideas for proving Theorem \ref{thm:sres} above.
	The three key steps for verifying the theorem are listed as follows.
	\begin{enumerate}[(1)]
		\item Convert the $\delta$th subresultant polynomial in roots from power to Newton basis (see Lemma \ref{lem:Sd_in_roots_NB}). The key in this step is the transition matrix between the power and Newton bases.
		\item  Convert the $\delta$th subresultant polynomial in Newton basis from a rational expression in roots to a determinant expression in coefficients (see Lemma \ref{lem:sres_in_det}). To achieve the goal, we use the companion matrix of a polynomial in Newton basis as the bridge to connect the expression in roots and that in coefficients together.

		\item Convert the resulting determinant expression in coefficients to its equivalent form of determinant polynomial as in Theorem \ref{thm:sres} (see Lemma \ref{lem:gen_det}).
	\end{enumerate}
	
	It is seen that Lemmas \ref{lem:Sd_in_roots_NB}, \ref{lem:sres_in_det} and \ref{lem:gen_det} are the key ingredients for proving the main theorem of the paper. Meanwhile,  they are interesting on their own. We hope that they can be useful for tackling other problems in the future.

		For the sake of simplicity, we introduce the following short-hands which will be frequently used in the proofs of the key lemmas.
		\begin{notation}\ 
			\begin{itemize}
				\item $\bar{x}_i=(x^{i-1},\ldots,x^{0})$;
				\item $\bar{x}_i(\alpha_j)=(\alpha_j^{i-1},\ldots,\alpha_j^{0})$;
				\item $\tilde{B}^{\lambda}_{i}=(B_{0},\ldots,B_{i-1})$;
				\item $\operatorname*{diag}F_i(\alpha)=\begin{bmatrix}
					F_i(\alpha_1)&&\\
					&\ddots&\\
					&&F_i(\alpha_{d_0})
				\end{bmatrix}$.
			\end{itemize}
		\end{notation}

	Lemma \ref{lem:Sd_in_roots_NB} presented below is targeted at fulfilling the goal in Step (1).
	\begin{lemma}\label{lem:Sd_in_roots_NB}
		
			We have
			\begin{equation}\label{eq:root_expression_newton}
				S_{\delta}(F)=(-1)^{\sigma'}a_{0d_0}^{\delta_0}\cdot\det M_{\delta}^{\lambda}\big/\det \tilde{B}^{\lambda}(\alpha)
			\end{equation}
			where
			\begin{itemize}
				\item $M_{\delta}^{\lambda}=\left[\begin{array}{lcl}
					(\tilde{B}^{\lambda}_{\delta_1}(\alpha_1))^TF_1(\alpha_1)&\cdots&(\tilde{B}^{\lambda}_{\delta_1}(\alpha_{d_0}))^TF_1(\alpha_{d_0})\\
					~~~~~~~~~~~~\vdots&&~~~~~~~~~~~~~\vdots\\
					(\tilde{B}^{\lambda}_{\delta_t}(\alpha_1))^TF_t(\alpha_1)&\cdots&(\tilde{B}^{\lambda}_{\delta_t}(\alpha_{d_0}))^TF_t(\alpha_{d_0})\\
					(\tilde{B}^{\lambda}_{\varepsilon}(\alpha_1))^T(x-\alpha_1)&\cdots&(\tilde{B}^{\lambda}_{\varepsilon}(\alpha_{d_0}))^T(x-\alpha_{d_0})
				\end{array}\right]$,
				\item $\tilde{B}^{\lambda}(\alpha)
				=\left[ {\begin{array}{*{20}{c}}
						{\tilde{B}^{\lambda}_{d_0}({\alpha _1})}\\
						\vdots \\
						{\tilde{B}^{\lambda}_{d_0}({\alpha _{d_0}})}
				\end{array}} \right]^T$ \item $\sigma'=\left(\sum_{i=1}^{t}\binom{\delta_i}{2}\right)+\binom{\varepsilon}{2}-\binom{d_0}{2}$, and
				\item $\delta_0$ and $\varepsilon$ are as in Definition \ref{def:sres}.
			\end{itemize}
		
	\end{lemma}
	
	\begin{remark}
		It is noted that the rational expression \eqref{eq:root_expression_newton} in Lemma \ref{lem:Sd_in_roots_NB} should be interpreted the same way as that for interpreting Equation \eqref{eq:Sk_AB_roots} and the rational expression \eqref{eqs:sres} for $S_{\delta}$ in
		Definition \ref{def:sres}. Otherwise, it would make no sense when $\alpha_i=\lambda_j$ for some $i,j$.
	\end{remark}
	\begin{proof}
		
			Recall $S_{\delta}(F)=a_{0d_0}^{\delta_0}\cdot\det M_{\delta}\big/\det V$. We will show 
			$$\det \tilde{B}^{\lambda}(\alpha) = (-1)^{\binom{d_0}{2}}\det V\quad\text{and}\quad \det M_{\delta}^{\lambda}=(-1)^{\sum_{i=1}^t\binom{\delta_i}{2}+\binom{\varepsilon}{2}}\det M_{\delta}$$ respectively, from which the lemma is followed.
			For this purpose,  we assume $U$ is
			the transition matrix from $\bar{x}_{d_0}$
			to $\tilde{B}^{\lambda}_{d_0}$, i.e.,
			$\tilde{B}^{\lambda}_{d_0} = \bar x_{d_0} \cdot U$. It is easy to see that  $U$ is a skew lower-triangular matrix with its anti-diagonal entries to be all ones.

		\smallskip
		(1) 
			Note that
			$$\left[ {\begin{array}{*{20}{c}}
					{\tilde{B}^{\lambda}_{d_0}({\alpha _1})}\\
					\vdots \\
					{\tilde{B}^{\lambda}_{d_0}(\alpha _{d_0})}
			\end{array}} \right] = \left[ {\begin{array}{*{20}{c}}
					{\bar x_{d_0}(\alpha_1)}\\
					\vdots \\
					{\bar x_{d_0}(\alpha_{d_0})}
			\end{array}} \right] \cdot U=V^T\cdot U$$
			Taking determinants on the left-hand and right-hand sides of the above equation and observing that $U$ is a skew-lower-triangle matrix with anti-diagonal entries to be all ones, we immediately get
			$$\det \tilde{B}^{\lambda}(\alpha)  = \det V^T \cdot \det U=(-1)^{\binom{d_0}{2}}  \det V$$
		
		\smallskip
		\indent (2) To show $\det M_{\delta}=(-1)^{\sum_{i=1}^t\binom{\delta_i}{2}+\binom{\varepsilon}{2}}\det M_{\delta}^{\lambda}$, we partition $M_{\delta}$ and $M_{\delta}^{\lambda}$ into $t+1$ parts, that is,
			\[M_{\delta}=
			\begin{bmatrix}
				M_1\\\vdots\\M_t\\X_\varepsilon
			\end{bmatrix}\quad\text{and}\quad M_{\delta}^{\lambda}=
			\begin{bmatrix}
				M_{1}^{\lambda}\\\vdots\\M_{t}^{\lambda}\\X_{\varepsilon,\lambda}
			\end{bmatrix}\]
			where
			\begin{align*}
				M_i&=\begin{bmatrix}
					\bar{x}_{\delta_i}(\alpha_1){F_i}(\alpha _1)&\  \cdots &{\bar{x}_{\delta_i}(\alpha_{d_0})F({\alpha _{{d_0}}})}
				\end{bmatrix}\\
				X_{\varepsilon}&=
				\begin{bmatrix}
					{\bar{x}_{\varepsilon}(\alpha_1)(x - {\alpha _1})}& \cdots &{\bar{x}_{\varepsilon}(\alpha_{d_0})(x - {\alpha _{{d_0}}})}
				\end{bmatrix}\\
				M_{i}^{\lambda}&=\begin{bmatrix}
					{(\tilde{B}^{\lambda}_{\delta_i}({\alpha _1}))^T{F_i}({\alpha _1})}& \ \, \cdots\  &{(\tilde{B}^{\lambda}_{\delta_i}({\alpha _{{d_0}}}))^T{F_i}({\alpha _{{d_0}}})}
				\end{bmatrix}\\
				X_{\varepsilon}^{\lambda}&=
				\begin{bmatrix}
					{(\tilde{B}^{\lambda}_{\varepsilon}({\alpha _1}))^T(x - {\alpha _1})}& \cdots &{(\tilde{B}^{\lambda}_{\varepsilon}({\alpha _{{d_0}}}))^T(x - {\alpha _{{d_0}}})}\\
				\end{bmatrix}
			\end{align*}
			Note that
			\begin{align*}
				M_{i}^{\lambda}
				&=\begin{bmatrix}
					{(\tilde{B}^{\lambda}_{\delta_i}({\alpha _1}))^T}& \cdots &{(\tilde{B}^{\lambda}_{\delta_i}({\alpha _{{d_0}}}))^T}
				\end{bmatrix}\cdot \operatorname*{diag}F_i(\alpha)\\
				&=\begin{bmatrix}I_{\delta_i}&0_{d_0-\delta_i}
				\end{bmatrix}\cdot\begin{bmatrix}
					{(\tilde{B}^{\lambda}_{d_0}({\alpha _1}))^T}& \cdots &{(\tilde{B}^{\lambda}_{d_{0}}({\alpha _{{d_0}}}))^T}
				\end{bmatrix}\cdot \operatorname*{diag}F_i(\alpha)\\
				&=\begin{bmatrix}
					I_{\delta_i}&0_{d_0-\delta_i}
				\end{bmatrix}\cdot U^T\cdot
				\begin{bmatrix}
					\bar x_{d_0}(\alpha_1) & \cdots &\bar x_{d_0}(\alpha_{d_0})
				\end{bmatrix}\cdot \operatorname*{diag}F_i(\alpha)
			\end{align*}
			where  $I_{\delta_i}$ is the identity matrix of order $\delta_i$ and  $0_{d_0-\delta_i}$ is the zero matrix of order $\delta_i\times{(d_0-\delta_i})$. Since $U$ is skew lower-triangular, so is $U^T$, which inspires us to partition $U^T$ in the following manner
			$$U^T=\begin{bmatrix}&U_{\delta_i}\\
				U_{\delta_i}'&\cdot\end{bmatrix}$$
			where $U_{\delta_i}$ and $U_{\delta_i}'$ are skew lower-triangular matrices of orders $\delta_i$ and $d_0-\delta_i$, respectively. Then \begin{align*}
				M_{i}^{\lambda}&=\begin{bmatrix}
					I_{\delta_i}&0_{d_0-\delta_i}
				\end{bmatrix}\cdot \begin{bmatrix}&U_{\delta_i}\\
					U_{\delta_i}'&\cdot\end{bmatrix}
				\cdot
				\begin{bmatrix}
					\bar x_{d_0}(\alpha_1) & \cdots &\bar x_{d_0}(\alpha_{d_0})
				\end{bmatrix}\cdot \operatorname*{diag}F_i(\alpha)\\
				&=U_{\delta_i}
				\cdot
				\begin{bmatrix}
					\bar x_{\delta_i}(\alpha_1) & \cdots &\bar x_{\delta_i}(\alpha_{d_0})
				\end{bmatrix}\cdot \operatorname*{diag}F_i(\alpha)\\
				&=U_{\delta_i}M_{i}
			\end{align*}
			With the similar technique, one can derive that $X_{\varepsilon}^{\lambda}=U_{\varepsilon}X_{\varepsilon}$. Assembling $M_{1}^{\lambda}$,$\ldots, M_{t}^{\lambda}$ and $X_{\varepsilon}^{\lambda}$ together, we have
			\[
			\begin{bmatrix}
				M_{1}^{\lambda}\\\vdots\\M_{t}^{\lambda}\\X_{\varepsilon}^{\lambda}
			\end{bmatrix}=
			\begin{bmatrix}
				U_{\delta_1}M_{1}\\\vdots\\U_{\delta_t}M_{t}\\U_{\varepsilon}X_{\varepsilon}
			\end{bmatrix}=
			\operatorname*{diag}\begin{bmatrix}
				U_{\delta_1}&\cdots&U_{\delta_t}&U_{\varepsilon}
			\end{bmatrix}\cdot
			\begin{bmatrix}
				M_{1}\\\vdots\\M_{t}\\X_{\varepsilon}
			\end{bmatrix}\]
			Taking the determinants on the left-hand and right-hand sides of the above equation and noting that $\det U_{k}=(-1)^{\binom{k}{2}}$, we immediately get
			$$\det M_{\delta}^{\lambda}=\left(\prod_{i=1}^t\det U_{\delta_i}\right)\cdot\det U_{\varepsilon}\cdot \det M_{\delta}=(-1)^{\sum_{i=1}^t\binom{\delta_i}{2}+\binom{\varepsilon}{2}}\det M_{\delta}$$ 
			The proof is completed.
	\end{proof}

	Next, we convert the subresultant polynomial in Newton basis from an expression in roots to that in coefficients. To achieve the goal, we introduce the following proposition, which shows that the companion matrix of a polynomial in Newton basis represents an endomorphism
	of $\mathbb{F}_{n-1}[x]$ defined by the multiplication by $cx$ for some $c$ with respect to the Newton basis (\citet[Proposition 5.1.8]{1996_Fuhrmann}), which captures the very essential property of companion matrices. Proposition \ref{the:general-companion} provides us with a bridge to connect the subresultant polynomial expression in roots and that in coefficients together.
	
	\begin{proposition}\label{the:general-companion}
		Given  $\lambda\in\mathbb{F}^{n}$, assume $P=\sum_{i=0}^np_iB_i$ where $B_i$'s are as in \eqref{def:nwbasis}.
		Let ${\Lambda _{{\lambda},P}}$ be the
		companion matrix of $P$ in $\tilde{B}^\lambda=(B_0,B_1,\ldots,B_{n-1})$. Then we have
		\begin{enumerate}[(1)]
			\item $p_nx \cdot \tilde{B}^\lambda{ \equiv _P}\tilde{B}^\lambda\cdot {\Lambda _{{\lambda},P}}$, and
			\item
			$ Q\cdot\tilde{B}^\lambda{ \equiv _P}\tilde{B}^\lambda \cdot Q({\Lambda _{\lambda,P}}/p_{n})$ for any $Q\in\mathbb{F}[x]$,
		\end{enumerate}
	where ${ \equiv _P}$ denotes the modulo operation of $P$.
	\end{proposition}
	
	With the help of Proposition \ref{the:general-companion}, we now prove the following lemma, which allows us to convert the subresultant polynomial in Newton basis from an expression in roots to that in coefficients.
	\begin{lemma}\label{lem:sres_in_det}
		Let ${{N}_{\lambda,\delta} }(F)$ be as in \eqref{eqs:N_lambda_delta}
			and
			\[X_{B^\lambda,\varepsilon} =
			\begin{bmatrix}
				x-\lambda_1&-1&&&&&&\\
				&x-\lambda_2&-1&&&&&\\
				&&\ddots&\ddots&&&&\\
				&&&x-\lambda_{\varepsilon}&-1&&&
			\end{bmatrix}_{\varepsilon\times d_0}\]
			where $\varepsilon=d_0-|\delta|$. Then
			\[S_{\delta}(F)=(-1)^{\sigma'}\cdot a_{0d_0}^{\delta_0}\cdot\det
			\begin{bmatrix}
				{{N}_{\lambda,\delta} }(F)\\X_{B^\lambda,\varepsilon}
			\end{bmatrix}_{d_0\times d_0}\]
			where $\sigma'=\left(\sum_{i=1}^{t}\binom{\delta_i}{2}\right)+\binom{\varepsilon}{2}-\binom{d_0}{2}$.
	\end{lemma}
	
	\begin{proof}
	Recall $M_{\delta}^{\lambda}$ in Lemma \ref{lem:Sd_in_roots_NB}.  
		We only need to show that
		\begin{equation}\label{eq:main_thm}
			\det M_{\delta}^{\lambda}=\det
			\begin{bmatrix}{{N}_{\lambda,\delta} }(F)\\X_{B^\lambda,\varepsilon}
			\end{bmatrix}\cdot\det \tilde{B}^\lambda(\alpha)=\det
			\begin{bmatrix}
				{{N}_{\lambda,\delta} }(F)\cdot \tilde{B}^\lambda(\alpha)\\
				X_{B^\lambda,\varepsilon}\cdot \tilde{B}^\lambda(\alpha)
			\end{bmatrix}
		\end{equation}
		This inspires us to partition $M_{\delta}^{\lambda}$  in the following way:
		$
		M_{\delta}^{\lambda}=\begin{bmatrix}D_1\\D_2\end{bmatrix}
		$
		where $D_1$ consists of the first $t$ blocks of $M_{\delta}^{\lambda}$ and $D_2$ is the block involving $x$. The remaining part will be dedicated to showing that
		\[D_1={{N}_{\lambda,\delta} }(F)\cdot \tilde{B}^\lambda(\alpha)\ \ \text{and} \ \ D_2=X_{B^\lambda,\varepsilon}\cdot \tilde{B}^\lambda(\alpha),\]
		respectively.
		
		Recall that
		$$
		{{{N}_{\lambda,\delta} }(F)} =
		\begin{bmatrix}
			{R_{11}}&\cdots&R_{1\delta_1}&\cdots&\cdots
			&{R_{t1}}&\cdots&R_{t\delta_t}
		\end{bmatrix}^T
		$$
		where $R_{ij}$ is the $j$th column of $F_i({\Lambda _{{\lambda,F_0}}}/a_{0d_0})$ and $a_{0d_0}$ is the leading coefficient of $F_0$. 
		The expansion of ${{N}_{\lambda,\delta} }(F)\cdot \tilde{B}^\lambda({\alpha})$ yields
			\[
			{{N}_{\lambda,\delta} }(F)\cdot \tilde{B}^\lambda({\alpha}) = \left[ {\begin{array}{*{20}{c}}
					\tilde{M}_{1}^{\lambda}\\
					\vdots\\
					\tilde{M}_{t}^{\lambda}
			\end{array}} \right]
			\]
			where
			\begin{equation}\label{eq:expansion_BN}
				\tilde{M}_{i}^{\lambda}=\begin{bmatrix}
					{\tilde{B}^\lambda({\alpha _1})R_{i1}}& \cdots &{\tilde{B}^\lambda({\alpha _{{d_0}}})R_{i1}}\\
					\vdots &{}& \vdots \\
					{\tilde{B}^\lambda({\alpha _1})R_{i{\delta _i}}}& \cdots &{\tilde{B}^\lambda({\alpha _{{d_0}}})R_{i{\delta _i}}}
				\end{bmatrix}_{\delta_i\times d_0}
			\end{equation}
			Denote $F_i({\Lambda _{{\lambda,F_0}}}/a_{0d_0})$ with $R_i$, i.e.,
			$R_i=
			\begin{bmatrix}
				R_{i1}&\cdots&R_{id_0}
			\end{bmatrix}$. By Proposition \ref{the:general-companion}-(2),
			\begin{equation}\label{eq:BQmodularF0}
				\tilde{B}^\lambda(x)\cdot F_i(x){ \equiv _{F_0}}\tilde{B}^\lambda(x)\cdot  R_i
			\end{equation}
			Substituting $x=\alpha_j$ into \eqref{eq:BQmodularF0} and noting that $F_0(\alpha_j)=0$, we get
			$$\tilde{B}^\lambda({\alpha _j}) \cdot {F_i}({\alpha _j}) = \tilde{B}^\lambda({\alpha _j})\cdot R_{i}$$
			Taking the $k$th columns of both sides of the above equation, we have
			$${B_{k- 1}}({\alpha _j}) \cdot {F_i}({\alpha _j}) = \tilde{B}^\lambda({\alpha _j}) \cdot R_{ik}$$
			The application of the above relation on \eqref{eq:expansion_BN} immediately yields  
			$$\tilde{M}_i^{\lambda}=\begin{bmatrix}
				{\tilde{B}^{\lambda}_{\delta_i}({\alpha _1}){F_i}({\alpha _1})}& \cdots &{\tilde{B}^{\lambda}_{\delta_i}({\alpha _{{d_0}}}){F_i}({\alpha _{{d_0}}})}
			\end{bmatrix}$$ which is exactly the $i$th block of $M_{\delta}^{\lambda}$ for $i=1,\ldots,t$. Hence
			\begin{equation}\label{eq:simplified_D1}
				{{N}_{\lambda,\delta} }(F)\cdot \tilde{B}^\lambda({\alpha}) =D_1
		\end{equation}
		
		Let $X_s$ be the $s$th row of $X_{B^\lambda,\varepsilon}$, i.e.,
		\begin{align*}
			{X_s} &= \left[ {\begin{array}{*{20}{c}}
					0& \cdots & 0&
					{x - {\lambda _s}} & { - 1}&
					0&\cdots & 0
			\end{array}} \right]\\[-5pt]
			&\hspace{6.5em}{\substack{\uparrow\\s\text{th column}}}
		\end{align*}
		where $s \in \{ 1,2, \ldots ,\varepsilon \} $.
		Simple calculation yields
			\[
			{X_s}\cdot \tilde{B}^\lambda({\alpha _i})^T = (x - {\lambda _s}) \cdot {B_{s - 1}}({\alpha _i}) - {B_s}({\alpha _i})= {B_{s - 1}}({\alpha _i})(x - {\alpha _i})
			\]
			It follows that $X_{B^\lambda,\varepsilon}\tilde{B}^\lambda({\alpha _i})^T={\tilde{B}^\lambda_{\varepsilon}}({\alpha _i})(x - {\alpha _i}).$
			Thus
			\begin{align}
				X_{B^\lambda,\varepsilon}\cdot \tilde{B}^\lambda({\alpha}) &=X_{B^\lambda,\varepsilon}\cdot\left[ {\begin{array}{*{20}{c}}
						{\tilde{B}^\lambda({\alpha _1})}^T&
						\cdots &
						{\tilde{B}^\lambda({\alpha _{{d_0}}})}^T
				\end{array}} \right]\notag\\
				&=\left[ {\begin{array}{*{20}{c}}
						{\tilde{B}^\lambda_{\varepsilon}}({\alpha _1})(x - {\alpha _1})&
						\cdots &
						{\tilde{B}^\lambda_{\varepsilon}}({\alpha _{d_0}})(x - {\alpha _{d_0}})
				\end{array}} \right]\label{eq:expansion_BX}
			\end{align}

			Combining \eqref{eq:simplified_D1} and \eqref{eq:expansion_BX}, we have
			$
			\begin{bmatrix}{{N}_{\lambda,\delta} }(F)\\X_{B^\lambda,\varepsilon}
			\end{bmatrix}\cdot \tilde{B}^\lambda(\alpha)
			=M_{\delta}^{\lambda}
			$.
			Applying Lemma \ref{lem:Sd_in_roots_NB},  we finally attain
			\[S_{\delta}(F)=(-1)^{\sigma'}\cdot a_{0d_0}^{\delta_0}\det M_{\delta}^{\lambda}/\det\tilde{B}^\lambda(\alpha)=(-1)^{\sigma'}\cdot a_{0d_0}^{\delta_0}\cdot\det\begin{bmatrix}{{N}_{\lambda,\delta} }(F)\\X_{B^\lambda,\varepsilon}
			\end{bmatrix}\]
			where $\sigma'=\left(\sum_{i=1}^{t}\binom{\delta_i}{2}\right)+\binom{\varepsilon}{2}-\binom{d_0}{2}$.
	\end{proof}
	
	\begin{remark}
	Lemma \ref{lem:sres_in_det} is a generalized version of \citet[Theorem 27-2]{hong2021subresultant}.
			The latter can be obtained by specializing $\lambda_1=\cdots=\lambda_{\varepsilon}=0$ in Lemma \ref{lem:sres_in_det}.
	\end{remark}
	
	The final step is to convert the generalized subresultant polynomial from a {determinant} expression to an equivalent determinant polynomial expression.
	For this purpose, we present a more general result than what we need. The more general result is presented in the hope that it would be useful for other related problems.

	\begin{lemma}\label{lem:gen_det}
		Given $$M= \left[ {\begin{array}{*{20}{c}}
				{{a_{1,1}}}& \cdots &{{a_{1,n }}}\\
				\vdots &{}& \vdots \\
				{{a_{n-k,1}}}& \cdots &{{a_{n-k,n}}}
		\end{array}} \right]_{(n-k)\times n}$$
		and
		$P=(P_0,P_1,\ldots,P_k)\subseteq\mathbb{F}[x]$,
		we have
		$$\operatorname*{detp}\nolimits_P M=(-1)^{(n-1)k}\cdot P_0\cdot\det \begin{bmatrix}M\\X_{P}\end{bmatrix},$$
		where
		\[X_{P}=\left[
		\begin{array}{cccccccc}
			P_1/P_0 & -1 & &&&&& \\
			& P_2/P_1 & -1 & &&&& \\
			&  & \ddots& \ddots&&&&\\
			&  && P_k/P_{k-1} & -1&& &
		\end{array}
		\right]_{k\times n}\]
	\end{lemma}
	
	\begin{proof}
		By Definition \ref{def:detp},
		\[ \operatorname*{detp}\nolimits_P M=\sum_{i=0}^{k}\det
		\left[ {\begin{array}{*{20}{c}}
				a_{1,i+1}&{{{a_{1,k+2}}} }& \cdots &{{{a_{1,n}}} }\\
				a_{2,i+1}&{{{a_{2,k+2}}} }&\cdots &{{a_{2,n}}}\\
				\vdots &{}& &\vdots \\
				a_{n-k,i+1}&{{{a_{n-k,k+2}}} }&\cdots &{{a_{n-k,n}}}
		\end{array}} \right]_{(n-k)\times(n-k)}\cdot {P_{i}}
		\]
		Making use of the multi-linearity of determinants, we have
		\begin{align*}
			\operatorname*{detp}\nolimits_P M&=\sum_{i=0}^{k}\det
			\left[ {\begin{array}{*{20}{c}}
					a_{1,i+1}P_{i}&{{{a_{1,k+2}}} }& \cdots &{{{a_{1,n}}} }\\
					a_{2,i+1}P_{i}&{{{a_{2,k+2}}} }&\cdots &{{a_{2,n}}}\\
					\vdots &{}& &\vdots \\
					a_{n-k,i+1}P_{i}&{{{a_{n-k,k+2}}} }&\cdots &{{a_{n-k,n}}}
			\end{array}} \right]_{(n-k)\times(n-k)}\\
			&=\det
			\left[ {\begin{array}{*{20}{c}}
					\sum_{i=0}^{k}a_{1,i+1}P_{i}&{{{a_{1,k+2}}} }& \cdots &{{{a_{1,n}}} }\\
					\sum_{i=0}^{k}a_{2,i+1}P_{i}&{{{a_{2,k+2}}} }&\cdots &{{a_{2,n}}}\\
					\vdots &{}& &\vdots \\
					\sum_{i=0}^{k}a_{n-k,i+1}P_{i}&{{{a_{n-k,k+2}}} }&\cdots &{{a_{n-k,n}}}
			\end{array}} \right]_{(n-k)\times(n-k)}
		\end{align*}
		The next step is tricky, but it is the key to relating the determinant polynomial with the desired determinant. We first insert $k$ columns, that is, the first $k+1$ columns of $M$ with the first column excluded, between the first two columns of the matrix on the right-hand side of the above equation and then
		add $k$ rows of the special shape as given in the following matrix. Then we have
		\[
		\operatorname*{detp}\nolimits_P M=(-1)^{\tau}\det
			\setlength{\arraycolsep}{1pt}
			\left[\begin{array}{c|cccc|ccc}
				\sum_{i=0}^{k}a_{1,i+1}P_{i}&a_{1,2}&a_{1,3}&\cdots&a_{1,k+1}&{{{a_{1,k+2}}} }& \cdots &{{{a_{1,n}}} }\\
				\sum_{i=0}^{k}a_{2,i+1}P_{i}&a_{2,2}&a_{2,3}&\cdots&a_{2,k+1}&{{{a_{2,k+2}}} }&\cdots &{{a_{2,n}}}\\
				\vdots &\vdots&&\vdots&\vdots&&\vdots \\
				\sum_{i=0}^{k}a_{n-k,i+1}P_{i}&a_{n-k,2}&a_{n-k,3}&\cdots&a_{n-k,k+1}&{{{a_{n-k,k+2}}} }&\cdots &{{a_{n-k,n}}}\\\hline
				& -1 & &&&&& \\
				& P_2/P_1 & -1 & &&&& \\
				& & \ddots& \ddots&&&&\\
				&  && P_k/P_{k-1} & -1&& &\\
			\end{array}
			\right]_{n\times n}\]
		where
		$$\tau=k+\sum_{i=1}^k(1+i)+\sum_{i=1}^k(n-k+i)\equiv (n-1)k\mod 2.$$
		The equation holds because there is only one non-zero minor in the last $k$ rows, which is bounded by the delimitations.
		
		By subtracting the $i$th column multiplied by $P_{i-1}$
		for $i=2,\ldots,k+1$ from the first column and then taking the factor $P_0$ from the first column, we obtain
		\[\operatorname*{detp}\nolimits_PM=(-1)^{(n-1)k}\cdot P_0\cdot\det
		\begin{bmatrix}
			M\\X_{P}
		\end{bmatrix}.\]
		The proof is completed.
	\end{proof}
	
	\begin{remark}
		By Lemma \ref{lem:gen_det}, the generalized subresultant polynomial developed in \citet{DANDREA2023} and \citet[Equation (7)]{2006D'Andrea} for two univariate polynomials corresponding to an arbitrary set of monomials can be rewritten into a single determinant formula, which may bring some convenience for analyzing the properties of the generalized subresultant polynomials. In the Appendix section, we provide a detailed derivation for the formula.
	\end{remark}
	
	Now we are ready to prove the main result of the paper.
	
	\begin{proof}[Proof of Theorem \ref{thm:sres}]
		By Lemma \ref{lem:sres_in_det}, 
			$$S_{\delta}(F)=(-1)^{\sigma'}\cdot a_{0d_0}^{\delta_0}\cdot\det
			\begin{bmatrix}
				{{N}_{\lambda,\delta} }(F)\\X_{B^\lambda,\varepsilon}
			\end{bmatrix}_{d_0\times d_0}$$
			where $\sigma'=\left(\sum_{i=1}^{t}\binom{\delta_i}{2}\right)+\binom{\varepsilon}{2}-\binom{d_0}{2}$.
		Next we specialize $M$ and $P$ in Lemma \ref{lem:gen_det} with ${{N}_{\lambda,\delta} }(F)$ and $(B_0,\ldots,B_{\varepsilon})$, respectively. Then $n$ and $k$ in Lemma \ref{lem:gen_det} are specialized with $d_0$ and $\varepsilon$ accordingly. Meanwhile, it is observed that
		$B_{i}/B_{i-1}=x-\lambda_i$ and $B_0=1$.
		Therefore, we can achieve the following
		\[S_{\delta}(F)=(-1)^{\sigma}\cdot a_{0d_0}^{\delta_0}\cdot\operatorname*{detp}\nolimits_{(B_0,\ldots,B_{\varepsilon})}{{N}_{\lambda,\delta} }(F)\]
			where $\sigma=\sigma'+(d_0-1)\varepsilon$.
		The proof is completed.
	\end{proof}
	
	\section{Application}\label{sec:application}

	The $\delta$th subresultant polynomial was originally proposed to solve the problem of parametric gcd for multiple polynomials. When combining  Theorems \ref{thm:icdeg} and \ref{thm:sres}, it can also be used to compute the gcd of numerical polynomials. In this section, we devise a method for computing the gcd of several univariate polynomials in Newton basis as an application of Theorem \ref{thm:sres}.

		The method relies on the following proposition.
	
	\begin{proposition}\label{prop:max_independant_set}
		
			Let $\theta={\rm icdeg} F$. Then the columns of the matrix $N_{\lambda,\theta}^T$, i.e.,
			\[{R_{11}},\ldots,R_{1\theta_1},\ldots
			,{R_{t1}},\ldots, R_{t\theta_t}
			\]
			form a maximal set of independent columns of the matrix
$$
	\begin{bmatrix}
		{F_1({\Lambda _{{\lambda,F_0}}}/a_{0d_0})}&\cdots&F_t({\Lambda _{{\lambda,F_0}}}/a_{0d_0})
\end{bmatrix}
$$
	\end{proposition}
	
	\begin{proof}
	
			We first show\ that  ${R_{11}},\ldots,R_{1\theta_1},\ldots
			,{R_{t1}},\ldots, R_{t\theta_t}$ are linearly independent. Otherwise,   ${N}_{\lambda,\theta}(F)$ is rank-deficient and thus $\operatorname*{detp}\nolimits_{(B_0,\ldots,B_{\varepsilon})} {N}_{\lambda,\theta}(F)=0$ where $\varepsilon=d_0-|\theta|$. By Theorem \ref{thm:sres},
			$$S_{\theta}=c\cdot \operatorname*{detp}\nolimits_{(B_0,\ldots,B_{\varepsilon})} {N}_{\lambda,\theta}(F)=0$$
			which contradicts with Theorem \ref{thm:icdeg}.

			Next we show that  $R_{ij}$ for $j>\theta_i$ can be written as a linear combination of ${R_{11}},\ldots,R_{1\theta_1},\ldots
			,$ ${R_{t1}},\ldots, R_{t\theta_t}$, which is equivalent to
			$$
			\operatorname*{detp}\nolimits_{(B_0,\ldots,B_{\varepsilon-1})}\,\begin{bmatrix}
				N_{\lambda,\theta}\\ 
				R_{ij}^T\end{bmatrix}=0
			$$
			Assume $R_{ij}=(R_{ij,1},\ldots,R_{ij,d_0})^T$. By the multi-linearity of determinant, 
			\begin{align}
				\operatorname*{detp}\nolimits_{(B_0,\ldots,B_{\varepsilon-1})}\begin{bmatrix}
					N_{\lambda,\theta}\\
					R_{ij}^T\end{bmatrix}=&\setlength{\arraycolsep}{2pt} 
				\det\begin{bmatrix}
					\sum^{\varepsilon-1}_{k=0}R_{11,k+1}\cdot B_k&R_{11,\varepsilon+1}&\cdots&R_{11,d_0}\\[-2pt]
					\vdots&\vdots&&\vdots\\
					\sum^{\varepsilon-1}_{k=0}R_{1\theta_1,k+1}\cdot B_k&R_{1\theta_1,\varepsilon+1}&\cdots&R_{1\theta_1,d_0}\\[-2pt]
					\vdots&\vdots&&\vdots\\
					\sum^{\varepsilon-1}_{k=0}R_{t1,k+1}\cdot B_k&R_{t1,\varepsilon+1}&\cdots&R_{t1,d_0}\\[-2pt]
					\vdots&\vdots&&\vdots\\
					\sum^{\varepsilon-1}_{k=0}R_{t\theta_t,k+1}\cdot B_k&R_{t\theta_t,\varepsilon+1}&\cdots&R_{t\theta_t,d_0}
					\\
					\sum^{\varepsilon-1}_{k=0}R_{ij,k+1}\cdot B_k&R_{ij,\varepsilon+1}&\cdots&R_{ij,d_0}
				\end{bmatrix}\notag\\
				=&\setlength{\arraycolsep}{2pt} 
				\det\begin{bmatrix}
					\sum^{d_0-1}_{k=0}R_{11,k+1}\cdot B_k&R_{11,\varepsilon+1}&\cdots&R_{11,d_0}\\[-2pt]
					\vdots&\vdots&&\vdots\\[-2pt]
					\sum^{d_0-1}_{k=0}R_{1\theta_1,k+1}\cdot B_k&R_{1\theta_1,\varepsilon+1}&\cdots&R_{1\theta_1,d_0}\\[-2pt]
					\vdots&\vdots&&\vdots\\[-2pt]
					\sum^{d_0-1}_{k=0}R_{t1,k+1}\cdot B_k&R_{t1,\varepsilon+1}&\cdots&R_{t1,d_0}\\[-2pt]
					\vdots&\vdots&&\vdots\\[-2pt]
					\sum^{d_0-1}_{k=0}R_{t\theta_t,k+1}\cdot B_k&R_{t\theta_t,\varepsilon+1}&\cdots&R_{t\theta_t,d_0}
					\\
					\sum^{d_0-1}_{k=0}R_{ij,k+1}\cdot B_k&R_{ij,\varepsilon+1}&\cdots&R_{ij,d_0}
				\end{bmatrix}\label{eqs:linear_dependence}
			\end{align}
			Note that $\sum^{d_0-1}_{k=0}R_{pq,k+1}\cdot B_k=\tilde{B}^{\lambda}\cdot R_{pq}$.
			By Proposition \ref{the:general-companion}-(2),
			$F_i\cdot\tilde{B}^\lambda{ \equiv _{F_0}}\tilde{B}^\lambda \cdot F_i({\Lambda _{\lambda,F_0}}/a_{0d_0})$. Comparing the $j$th elements of both sides,
			we obtain 
			$\tilde{B}^{\lambda}\cdot R_{pq}\equiv_{F_0}F_pB_{q-1}$, which implies that 
			\[ \tilde{B}^{\lambda}\cdot R_{pq}=F_pB_{q-1}+C_{pq}F_0\]
			The substitution of the above relation into \eqref{eqs:linear_dependence} yields
			\[\operatorname*{detp}\nolimits_{(B_0,\ldots,B_{\varepsilon-1})}\,\begin{bmatrix}
				N_{\lambda,\theta}\\
				R_{ij}^T\end{bmatrix}
			=
			\det\begin{bmatrix}
				F_1B_{0}+C_{11}F_0&R_{11,\varepsilon+1}&\cdots&R_{11,d_0}\\[-2pt]
				\vdots&\vdots&&\vdots\\[-2pt]
				F_1B_{\theta_1-1}+C_{1\theta_1}F_0&R_{1\theta_1,\varepsilon+1}&\cdots&R_{1\theta_1,d_0}\\[-2pt]
				\vdots&\vdots&&\vdots\\[-2pt]
				F_tB_{0}+C_{t1}F_0&R_{t1,\varepsilon+1}&\cdots&R_{t1,d_0}\\[-2pt]
				\vdots&\vdots&&\vdots\\[-2pt]
				F_tB_{\theta_t-1}+C_{t\theta_t}F_0&R_{t\theta_t,\varepsilon+1}&\cdots&R_{t\theta_t,d_0}
				\\
				F_iB_{\theta_i-1}+C_{ij}F_0&R_{ij,\varepsilon+1}&\cdots&R_{ij,d_0}\end{bmatrix}\]
			which indicates that $\operatorname*{detp}\nolimits_{(B_0,\ldots,B_{\varepsilon-1})}\begin{bmatrix}
				N_{\lambda,\theta}\\
				R_{ij}^T\end{bmatrix}
			$ belongs to the ideal generated by $F$. However, 
			$$\deg\left(\operatorname*{detp}\nolimits_{(B_0,\ldots,B_{\varepsilon-1})}\begin{bmatrix}
				N_{\lambda,\theta}\\
				R_{ij}^T\end{bmatrix}\right)
			\le\varepsilon-1=\deg \gcd F-1<\deg \gcd F$$
			Therefore, \[\operatorname*{detp}\nolimits_{(B_0,\ldots,B_{\varepsilon-1})}\begin{bmatrix}
				N_{\lambda,\theta}\\
				R_{ij}^T\end{bmatrix}
			=0\]
			Hence $R_{ij}^T$ is a linear combination of the rows in 
			$N_{\lambda,\theta}$. It immediately follows that ${R_{11}},\ldots,R_{1\theta_1},$
			$
			\ldots,{R_{t1}},\ldots, R_{t\theta_t}$ form a maximal set of independent columns of the matrix
			$\left[
				{F_1({\Lambda _{{\lambda,F_0}}}/a_{0d_0})}\ \ \cdots\ \ \right.$ $\left.F_t({\Lambda _{{\lambda,F_0}}}/a_{0d_0})
			\right]$.
	\end{proof}

	Now we describe the method for computing the gcd of several polynomials in Newton basis, where $F$ is assumed to be as in Notation \ref{notation} and expressed in the Newton basis $B^{\lambda}(x)$.
	\begin{enumerate}[(S1)]
		\item Determine the independent columns of ${F_1({\Lambda _{{\lambda,F_0}}}/a_{0d_0})}$ incrementally. More explicitly, add one column each time to test its linear dependency with the previous columns until we identify the first column which is linearly dependent with the previous columns. If the column is the $j$th column of ${F_1({\Lambda _{{\lambda,F_0}}}/a_{0d_0})}$, let $\theta_1=j-1$  and go to (S2).
		\item For $i=2,\ldots,t$, determine the columns of ${F_i({\Lambda _{{\lambda,F_0}}}/a_{0d_0})}$ linearly independent with  ${R_{11}},\ldots,$
        $R_{1\theta_1},\ldots,
	{R_{i1}},\ldots,R_{i\theta_i}$ in an incremental way. More explicitly, add one column of ${F_i({\Lambda _{{\lambda,F_0}}}/a_{0d_0})}$ each time to test its linear dependency with the previous columns 
			until we identify the first column which is linearly dependent with the previous columns. If the column is the $j$th column of ${F_i({\Lambda _{{\lambda,F_0}}}/a_{0d_0})}$, then let $\theta_i=j-1$  and go to the next loop.
		
		\item Construct $$N_{\lambda,\theta}=\begin{bmatrix}R_{11}&\cdots&R_{1\theta_1}&\cdots&
				\cdots&{R_{t1}}&\cdots& R_{t\theta_t}\end{bmatrix}^T$$ 
			Let $\varepsilon=d_0-|\theta|$. Then the gcd of $F$ in Newton basis is $\operatorname*{detp}\nolimits_{(B_0,\ldots,B_{\varepsilon})}N_{\lambda,\theta}$.
	\end{enumerate}

		We discuss the correctness of the above method below. 
	\begin{enumerate}[(1)]
		\item 
		Apply Proposition \ref{prop:max_independant_set} to $(F_0,F_1 )$. Then we have that a maximal set of linearly independent columns of ${F_1({\Lambda _{{\lambda,F_0}}}/a_{0d_0})}$ is ${R_{11}},\ldots,R_{1\theta_1}$ where $\theta_1=\deg F_0-\deg\gcd (F_0,F_1)$. So the first $j$ such that the $j$th column is linearly dependent with the previous $j-1$ columns must satisfy that $j=\theta_1+1$. In other words, $\theta_1=j-1$.
		
		\item Assume we have a maximal set of linearly independent columns of 
			${F_1({\Lambda _{{\lambda,F_0}}}/a_{0d_0})},$ $\ldots$, ${F_{i-1}({\Lambda _{{\lambda,F_0}}}/a_{0d_0})}$, say ${R_{11}},\ldots,R_{1\theta_1},\cdots
			{R_{i1}},\ldots,R_{(i-1)\theta_{i-1}}$ where $$\theta_j=\deg\gcd (F_0,\ldots,F_{j-1})-\deg\gcd (F_0,\ldots,F_{j})$$ It is noted that adding any other $R_{jk}$ with $j<i$ and $k>\theta_j$ will destroy the linearly independence. Now we add the columns of ${F_i({\Lambda _{{\lambda,F_0}}}/a_{0d_0})}$ into the linearly independent set one by one until we identify the first column which is linearly dependent with the previous columns.
			If the column is the $j$th column of ${F_i({\Lambda _{{\lambda,F_0}}}/a_{0d_0})}$, by Proposition \ref{prop:max_independant_set}, $j=\theta_i+1$ and thus $\theta_i=j-1$. 
		\item By Theorem \ref{thm:icdeg}, $\gcd F=S_{\theta}(F)$; by Theorem \ref{thm:sres}, $S_{\theta}(F)=c\cdot\operatorname*{detp}_{(B_0,\ldots,B_{\varepsilon})}N_{\lambda,\theta} $. Hence $\operatorname*{detp}\nolimits_{(B_0,\ldots,B_{\varepsilon})}N_{\lambda,\theta}$ is the gcd of $F$.
	\end{enumerate}

	\begin{example}[Continued from Example \ref{ex:sres_multi_polys}]
		
			We show $S_{(1,1)}$ obtained in Example  \ref{ex:sres_multi_polys} is exactly the gcd of $F$. Recall that 
			\[F_1(\Lambda_{\lambda,F_0}/4)=\left[ \begin {array}{ccc} \ \ 0&\ \ 0&\ \ 0\\ [2pt]
			-{\dfrac{3}{2}}&{
				\ \ \dfrac{3}{4}}&\ \ {\dfrac{9}{8}}\\[7pt]
			\ \ 1&-{\dfrac{1}{2}}&-{\dfrac{3}{4}}
			\end {array} \right]\quad\text{and}\quad F_2(\Lambda_{\lambda,F_0}/4)=
			\left[ \begin {array}{ccc}
			\ \ 2&\ \ 0&\ \ 0\\[2pt]
			-{\dfrac{7}{2}}&-
			{\dfrac{9}{4}}&-{\dfrac{27}{8}}\\[7pt]
			\ \ 5&\ \ {\dfrac{3}{2}}&{
				\ \ \dfrac{9}{4}}\end {array} \right]\]
			It is noted that there is only one linearly independent column in $F_1(\Lambda_{\lambda,F_0}/4)$ which is  the first column $R_{11}$. Thus $\theta_1=1$. With further calculation, we find that the first column $R_{21}$ of $F_2(\Lambda_{\lambda,F_0}/4)$ is linearly independent with $R_{11}$; however, the second column $R_{22}$ is linearly dependent with $(R_{11},R_{21})$. Thus $\theta_2=1$ and  $\theta=(1,1)$. We construct  $N_{\lambda,\theta}(F)$ as follows
			\[{N}_{\lambda,\theta}(F) =
			\begin{bmatrix}
				\ \ 0&\ \ 2\\[2pt]
				-{\dfrac{3}{2}}&{
					-\dfrac{7}{2}}\\[7pt]
				\ \ 1&\ \ 5
			\end{bmatrix}^T=
			\begin{bmatrix}
				0&-\dfrac{3}{2}&1\\[7pt]
				2&-\dfrac{7}{2}&5
			\end{bmatrix}\]
			Then its determinant polynomial ${\rm detp}_{(B_0,B_1)}{N}_{\lambda,\theta}(F)=-4B_1-2B_{0}$
			provides an expression for the gcd of $F$ and  $S_{\theta}$  only differs from it by a constant factor.
		
	\end{example}

	\section{Conclusion}\label{sec:conclusion}

		In this paper, we propose a method for formulating  subresultant polynomials of multiple univariate polynomials in Newton basis where the resulting subresultants polynomials are also expressed in the given Newton basis. The formulation does not require basis transformation and thus can avoid the numerical instability problem caused by basis change. To achieve this goal, we extend the determinant polynomial in power basis to that in Newton basis. By making use of
		the companion matrix in Newton basis, we
		construct a Barnett-type matrix  whose determinant polynomial in Newton basis yields the subresultant polynomial of multiple polynomials in Newton basis.

		It is noted that
		one may have infinite many choices for Newton interpolating nodes   $\lambda$. For each choice of   $\lambda$, we can write the input polynomials in the corresponding Newton basis and formulate the subresultant polynomial in the same basis. Therefore,  one can express a subresultant polynomial as infinitely many determinant forms, depending on which Newton basis is used. Then a natural question arises: does there exist
		$\lambda$'s such that the constructed subresultant matrix is interesting? For example, is there a particular way to choose the value of $\lambda$ for a specific polynomial set such that the matrix $N_{\lambda,\delta}$ exhibits nice structures? Another related question worthy of further investigation is how to compute the matrix $N_{\lambda,\delta}$ and its determinant polynomial in an efficient way.

	\medskip\noindent\textbf{Acknowledgements.} The authors would like to thank the anonymous reviewers for their helpful comments and insightful suggestions. The authors' work was
	supported by the National Natural Science Foundation of China (Grant No. 12261010), the Natural Science Foundation of Guangxi (Grant No.  2023GXNSFBA02\-6019), and the Natural Science Cultivation Project of GXMZU  (Grant No. 2022MDKJ001).
	
	\section*{Appendix}
	
	The Appendix is devoted to deriving a determinant formula for the generalized subresultant polynomial defined in \citet{2006D'Andrea} and \citet{DANDREA2023}.
	
	Consider $A,B\in\mathbb{F}[x]$ of the following form
	
		\[
		A=a_nx^n+\cdots+a_0,\quad B=b_mx^m+\cdots+b_0,\]
		where $a_nb_m\ne0.$
		For $P=(x^{\gamma_0},x^{\gamma_1},\ldots,x^{\gamma_k})\subseteq\mathbb{F}_{\ell}[x]$
		where $m\le \ell\le m+n-1$ and $k=m-\max(0,\ell-n+1)$, without loss of generality, we assume $0\le\gamma_0<\gamma_1<\ldots<\gamma_k$.
	Then  the subresultant matrix for $A$ and $B$ of order $\ell+1$ is constructed as follows
	\[M=\begin{bmatrix}
			a_0&\cdots&a_n\\
			&\ddots&&\ddots\\
			&&a_0&\cdots&a_n\\
			b_0&\cdots&b_m\\
			&\ddots&&\ddots\\
			&&b_0&\cdots&b_m\\
		\end{bmatrix}
		\hspace{-1.8em}\begin{array}{l}
			\left.
			\begin{array}{c}
				\\
				\\
				\\
			\end{array}
			\right\}~\max(0,\ell-n+1) \text{~rows}\\[20pt]
			\left.
			\begin{array}{c}
				\\
				\\
				\\
			\end{array}
			\right\}~\ell-m+1\text{~rows}\end{array}.
	\]
	which has the size $\left(\max(0,\ell-n+1)+(\ell-m+1)\right)\times(\ell+1)$.
	
	For the sake of simplicity, we partition $M$ by columns, that is,
	\[M=\begin{bmatrix}
		M_0&M_1&\cdots&M_{\ell}
	\end{bmatrix}.\]
	By \citet{DANDREA2023} and \citet[Equation (7)]{2006D'Andrea}, the generalized subresultant polynomial of $A$ and $B$ with respect to $P$ is
	$$s(x)=\sum_{j=0}^k(-1)^{\sigma_{j}}\det \widehat{M_{j}}\cdot x^{\gamma_j}$$
	where $\sigma_j=\sum_{\substack{0\le i\le k\\i\ne j}}{(i+\gamma_i)}$ and
	$\widehat{M_{j}}$ is the submatrix of $M$ obtained by deleting the columns $M_{\gamma_0}$, $M_{\gamma_1}$, $\ldots$, $M_{\gamma_k}$ but keeping $M_{\gamma_j}$. In order to write $s(x)$ in the form of a polynomial determinant, we need to permute the columns of
	$\widehat{M_{j}}$ so that  $M_{\gamma_j}$ becomes the first column. Denote the resulting matrix with $\widetilde{M_{j}}$. Then $\det \widehat{M_{j}}=(-1)^{j+\gamma_j}\det \widetilde{M_{j}}$ and hence
	\begin{align*}
		s(x)=&\sum_{j=0}^k(-1)^{\sigma_{j}+j+\gamma_j}\det \widetilde{M_{j}}\cdot x^{\gamma_j}\\
		=&\sum_{j=0}^k(-1)^{\sum\limits_{0\le i\le k}(i+\gamma_i)}\det \widetilde{M_{j}}\cdot x^{\gamma_j}\\
		=&(-1)^{\sum\limits_{0\le i\le k}(i+\gamma_i)}\sum_{j=0}^k\det \widetilde{M_{j}}\cdot x^{\gamma_j}\\
		=&(-1)^{\sum\limits_{0\le i\le k}(i+\gamma_i)}\operatorname{detp}_{P}\begin{bmatrix}
			M_{\gamma_0}&\cdots&M_{\gamma_k}&M_0&M_1&\cdots&M_{\gamma_0-1}&M_{\gamma_0+1}&\cdots
		\end{bmatrix}
	\end{align*}
	
	By Lemma \ref{lem:gen_det}, $s(x)$ can be written as
	\begin{align*}
		s(x)=&(-1)^{\sum\limits_{0\le i\le k}(i+\gamma_i)+\ell k}\cdot x^{\gamma_0}\\
		&\cdot\det\begin{bmatrix}
			M_{\gamma_0}&\cdots&M_{\gamma_k}&M_0&M_1&\cdots&M_{\gamma_0-1}&M_{\gamma_0+1}&\cdots\\
			x^{\gamma_1-\gamma_0}&-1\\
			&\ddots&\ddots\\
			&&x^{\gamma_k-\gamma_{k-1}}&-1
		\end{bmatrix}
	\end{align*}
	which gives us a determinant representation for the generalized subresultant polynomial $s(x)$.

\begin{thebibliography}{00}

\bibitem[Amiraslani et al.(2006)]{aadr2006}
A. Amiraslani, D. A. Aruliah, and R. M. Corless.
\newblock The {R}ayleigh quotient iteration for generalized companion matrix
pencils.
\newblock {\em Preprint}, 2006.

\bibitem[Arnon et al.(1984)]{arnon1984}
D. S. Arnon, G. E. Collins, and S. McCallum.
\newblock Cylindrical algebraic decomposition {I}: The basic algorithm.
\newblock {\em SIAM Journal on Computing}, 13(4):865--877, 1984.

\bibitem[Aruliah et al.(2007)]{acgs2007}
D. A. Aruliah, R. M. Corless, L. Gonzalez-Vega, and A. Shakoori.
\newblock Geometric applications of the {B}ezout matrix in the {L}agrange
basis.
\newblock In {\em Proceedings of the 2007 International Workshop on
	Symbolic-Numeric Computation}, SNC '07, pages 55–64, 2007.

\bibitem[Barnett(1969)]{barnett1969}
S. Barnett.
\newblock Degrees of greatest common divisors of invariant factors of two regular polynomial matrices.
\newblock In {\em Mathematical Proceedings of the Cambridge Philosophical
	Society}, volume~66, pages 241--245. Cambridge University Press, 1969.

\bibitem[Barnett(1970)]{barnett1970}
S. Barnett.
\newblock Degrees of greatest common divisors of invariant factors of two regular polynomial matrices.
\newblock {\em Linear Algebra and its Applications},  3(1):7--9, 1970.

\bibitem[Barnett(1971)]{barnett1971greatest}
S. Barnett.
\newblock Greatest common divisor of several polynomials.
\newblock In {\em Mathematical Proceedings of the Cambridge Philosophical
	Society}, volume~70, pages 263--268. Cambridge University Press, 1971.

\bibitem[Barnett(1983)]{barnett1983}
S. Barnett.
\newblock {\em Polynomials and Linear Control Systems}.
\newblock Marcel Dekker, Inc., USA, 1983.

\bibitem[Bini \& Gemignani(2004)]{bl2004}
D. A. Bini and L. Gemignani.
\newblock Bernstein-{B}ezoutian matrices.
\newblock {\em Theoretical Computer Science}, 315(2-3):319--333, 2004.

\bibitem[Bostan et al.(2017)]{bostan2017}
A. Bostan, C. D'Andrea, T. Krick, A. Szanto, and M. Valdettaro.
\newblock Subresultants in multiple roots: an extremal case.
\newblock {\em Linear Algebra and its Applications}, 529:185--198, 2017.

\bibitem[Carnicer \& Pe{\~n}a(1993)]{cp1993}
J. M. Carnicer and J. M. Pe{\~n}a.
\newblock Shape preserving representations and optimality of the {B}ernstein
basis.
\newblock {\em Advances in Computational Mathematics}, 1(2):173--196, 1993.

\bibitem[Cheng \& Labahn(2006)]{cl2006}
H. Cheng and G. Labahn.
\newblock On computing polynomial GCD in alternate bases.
\newblock In {\em Proceedings of the International Symposium on Symbolic and Algebraic Computation}, ISSAC '06, pages 47--54, 2006.

\bibitem[Collins(1967)]{collins1967}
G. E. Collins
\newblock Subresultants and reduced polynomial remainder sequences.
\newblock {\em Journal of the ACM (JACM)}, 14(1):128--142, 1967.

\bibitem[Collins \& Hong(1991)]{collins1991}
G. E. Collins and H. Hong. 
\newblock Partial cylindrical algebraic decomposition for quantifier elimination.
\newblock {\em Journal of Symbolic Computation}, 12(3):299--328,
1991.

\bibitem[Corless \& Litt(2001)]{2001_Corless_Litt}
R. M. Corless and G. Litt.
\newblock Generalized companion matrices for polynomials not expressed in monomial bases.
\newblock {\em Preprint}, 2001.

\bibitem[Cox \& D'Andrea(2021)]{cox2021}
D.~A. Cox and C. D'Andrea.
\newblock Subresultants and the {S}hape {L}emma.
\newblock {\em Mathematics of Computation}, 92:2355--2379, 2021.

\bibitem[D'Andrea \& Krick(2023)]{DANDREA2023}
C. D'Andrea and T. Krick.
\newblock Corrigendum to  ``{M}ultivariate subresultants in roots" [{J}. {A}lgebra {A}ppl. 302(1) (2006) 16–36].
\newblock {\em Journal of Algebra}, 635:838--839, 2023.

\bibitem[D'Andrea et al.(2006)]{2006D'Andrea}
C. D'Andrea, T. Krick, and A. Szanto.
\newblock Multivariate subresultants in roots.
\newblock {\em Journal of Algebra}, 302(1):16--36, 2006.

\bibitem[Delgado \& Pe{\~n}a(2003)]{jj2003}
J. Delgado and J. M. Pe{\~n}a.
\newblock A shape preserving representation with an evaluation algorithm of
linear complexity.
\newblock {\em Computer Aided Geometric Design}, 20(1):1--10, 2003.

\bibitem[Diaz-Toca \& Gonzalez-Vega(2002)]{diaz2002}
G. M. Diaz-Toca and L. Gonzalez-Vega.
\newblock Barnett's theorems about the greatest common
divisor of several univariate polynomials through {B}ezout-like matrices.
\newblock {\em Journal of Symbolic Computation}, 34(1):59--81, 2002.

\bibitem[Diaz-Toca \& Gonzalez-Vega(2004)]{diaz2004various}
G.~M. Diaz-Toca and L. Gonzalez-Vega.
\newblock Various new expressions for subresultants and their applications.
\newblock {\em Applicable Algebra in Engineering, Communication and Computing},
15(3):233--266, 2004.

\bibitem[Farouki \& Rajan(1987)]{fr1987}
R. T. Farouki and V. T. Rajan.
\newblock On the numerical condition of polynomials in {B}ernstein form.
\newblock {\em Computer Aided Geometric Design}, 4(3):191--216, 1987.

\bibitem[Fuhrmann(1996)]{1996_Fuhrmann}
P. A. Fuhrmann.
\newblock {\em A Polynomial Approach to Linear Algebra.}
\newblock Springer-Verlag, Berlin, Heidelberg, 1996.

\bibitem[Gonzalez-Vega et al.(1998)]{1998_Gonzalez_Recio_Lombardi}
L. Gonzalez-Vega, T. Recio, H. Lombardi, and M.-F. Roy.
\newblock {{S}turm-{H}abicht Sequences, Determinants and Real Roots of Univariate Polynomials}.
\newblock In {\em Quantifier Elimination and Cylindrical
	Algebraic Decomposition. Texts and Monographs in Symbolic Computation (A
	Series of the Research Institute for Symbolic Computation,
	Johannes-Kepler-University, Linz, Austria)}, pages 300--316. Springer, 1998.

\bibitem[Goodman \& Said(1991)]{goodman1991}
T. N. T. Goodman and H. B. Said.
\newblock Shape preserving properties of the generalised {B}all basis.
\newblock {\em Computer Aided Geometric Design}, 8(2):115--121, 1991.

\bibitem[Ho(1989)]{h1989}
C.-J. Ho.
\newblock {\em Topics in Algebraic Computing: Subresultants, GCD, Factoring and
	Primary Ideal Decomposition}.
\newblock PhD thesis, USA, 1989.

\bibitem[Hong(1999)]{1999_Hong}
H. Hong.
\newblock Subresultants in Roots.
\newblock Technical report, Department of Mathematics. North Carolina State University, 1999.

\bibitem[Hong(2002)]{hoon2002}
H. Hong.
\newblock Subresultants in roots.
\newblock In {\em Proceedings of 8th International Conference on Applications
	of Computer Algebra (ACA'2002)}. Volos, Greece, 2002.


\bibitem[Hong \& Yang(2021)]{hong2021subresultant}
H. Hong and J. Yang. 
\newblock Subresultant of several univariate polynomials.
\newblock {\em arXiv:2112.15370}, 2021.

\bibitem[Hong \& Yang(2023)]{2023_Hong_Yang}
H. Hong and J. Yang. 
\newblock Computing greatest common divisor of several parametric univariate polynomials via generalized subresultant polynomials.
\newblock {\em arXiv:2401.00408}, 2023.

\bibitem[Hou \& Wang(2000)]{houwang2000}
X.~R. Hou \& D.~M. Wang.
\newblock Subresultants with the {B}{\'e}zout matrix.
\newblock In {\em Computer Mathematics}, pages 19--28. World Scientific, 2000.

\bibitem[Imbach et al.(2017)]{2017_Imbach_Moroz_Pouget}
R. Imbach, G. Moroz, and M. Pouget.
\newblock A certified numerical algorithm for the topology of resultant and discriminant curves.
\newblock {\em Journal of Symbolic Computation},  80:285--306, 2017.

\bibitem[Kapur et al.(1994)]{kapur1994}
D. Kapur, T. Saxena, and L. Yang.
\newblock Algebraic and geometric reasoning using {D}ixon resultants.
\newblock In {\em Proceedings of the International Symposium on Symbolic and Algebraic Computation}, ISSAC '94, pages 99--107, 1994.

\bibitem[Lascoux \& Pragacz(2003)]{lascoux2003}
A. Lascoux and P. Pragacz.
\newblock Double {S}ylvester sums for subresultants and multi-{S}chur
functions.
\newblock {\em Journal of Symbolic Computation}, 35(6):689--710, 2003.

\bibitem[Li(2006)]{li2006}
Y. B. Li.
\newblock A new approach for constructing subresultants.
\newblock {\em Applied Mathematics and Computation}, 183(1):471--476, 2006.

\bibitem[Marco \& Mart{\'\i}nez(2007)]{mm2007}
A. Marco and J.-J. Mart{\'\i}nez.
\newblock {B}ernstein--{B}ezoutian matrices and curve implicitization.
\newblock {\em Theoretical computer science}, 377(1-3):65--72, 2007.

\bibitem[Mishra(1993)]{1993_Mishra}
B. Mishra.
\newblock {\em Algorithmic Algebra.}
\newblock Springer-Verlag, Berlin, Heidelberg, 1993.

\bibitem[Perrucci \& Roy(2018)]{2018_Perrucci_Roy}
D. Perrucci and M.-F. Roy.
\newblock A new general formula for the {C}auchy index on an interval with subresultants.
\newblock {\em Journal of Symbolic Computation}, 109:465--481, 2018.

\bibitem[Roy \& Szpirglas(2020)]{2020_Roy_Szpirglas}
M.-F. Roy and A. Szpirglas.
\newblock Sylvester double sums, subresultants and symmetric multivariate {H}ermite interpolation.
\newblock {\em Journal of Symbolic Computation}, 96:85--107, 2020.

\bibitem[Sylvester(1853)]{sylvester1853}
J. Sylvester.
\newblock On a theory of syzygetic relations of two rational integral
functions, comprising an application to the theory of {S}turm's functions,
and that of the greatest algebraic common measure.
\newblock {\em Phil. Trans}, 143:407--548, 1853.

\bibitem[Szanto(2008)]{2008_Szanto}
A. Szanto.
\newblock Solving over-determined systems by subresultant methods (with an appendix by {M}arc {C}hardin).
\newblock {\em Journal of Symbolic Computation}, 43(1):46--74, 2008.

\bibitem[Terui(2008)]{terui2008}
A. Terui.
\newblock Recursive polynomial remainder sequence and its subresultants.
\newblock {\em Journal of Algebra}, 320(2):633--659, 2008.

\bibitem[Wang(1998)]{wang1998}
D. M. Wang.
\newblock Decomposing polynomial systems into simple systems.
\newblock {\em Journal of Symbolic Computation}, 25(3):295--314, 1998.

\bibitem[Wang(2000)]{wang2000}
D. M. Wang.
\newblock Computing triangular systems and regular systems.
\newblock {\em Journal of Symbolic Computation}, 30(2):221--236, 2000.

\end{thebibliography}



\end{document}